%% file: paper.tex
\pdfoutput=1
\documentclass[conference]{IEEEtran}

\usepackage{epsfig,endnotes}
\usepackage{sepnum}
\usepackage{fancyvrb}
\usepackage{amsmath}
\usepackage{amssymb}
\usepackage{algorithmic}
\usepackage{color}
\usepackage{graphicx}
\usepackage{xspace}
\usepackage{booktabs}
\usepackage{bigstrut}
\usepackage{ifpdf}
\usepackage{url}
\usepackage{fixltx2e}
\usepackage{wrapfig}
\usepackage{moreverb}
\usepackage{amsthm}
\usepackage[bottom]{footmisc}
\usepackage{pifont}
\usepackage{placeins}
\usepackage{etoolbox}
\preto\subequations{\ifhmode\unskip\fi}

\usepackage{mfirstuc}

\usepackage[utf8]{inputenc}
\usepackage[T1]{fontenc}
\usepackage{ae,aecompl}

\usepackage{bookmark} 
\usepackage[font=scriptsize,labelfont=bf]{subcaption}
\usepackage[font=scriptsize,labelfont=bf]{caption}
\usepackage{paralist}
\usepackage{multirow}
\usepackage{todonotes}

\usepackage{tikz}
\usetikzlibrary{fpu}
\usetikzlibrary{arrows,shapes.geometric,positioning}

\newcommand{\empirical}[1]{{#1}}

\newcommand{\yearsData}{\empirical{9}\xspace}
\newcommand{\overlapFpReductonPercentage}{\empirical{20.13\%}\xspace}
\newcommand{\overlapTpReductonPercentage}{\empirical{11.96\%}\xspace}
\newcommand{\overlapSolvedDaysPercentage}{\empirical{98\%}\xspace}
\newcommand{\overlapFailedDaysPercentage}{\empirical{1.5\%}\xspace}
\newcommand{\overlapSolvedDaysTime}{\empirical{2.5}\xspace}
\newcommand{\numTotalSignatures}{\empirical{40,884}\xspace}
\newcommand{\numTotalCVEs}{\empirical{176}\xspace}

\newcommand{\etal}{{\em et al.}\xspace}

\def\tunable{tunable\xspace}
\def\Tunable{Tunable\xspace}

\newcommand{\squeezeup}{}

\begin{document}
\def\UrlFont{\footnotesize\ttfamily}
\sloppy

\title{Security: Doing Whatever is Needed... \\and Not a Thing More!}

\author{\IEEEauthorblockN{Omer Katz}
\IEEEauthorblockA{Technion\\
omerkatz@cs.technion.ac.il}
\and
\IEEEauthorblockN{Benjamin Livshits}
\IEEEauthorblockA{Imperial College London\\
b.livshits@imperial.ac.uk}
}

\setlength{\abovecaptionskip}{2mm}

\input{mymacros}

\maketitle

\input{abstract}
\input{intro}
\input{background}
\input{sampling}

\input{eval}
\input{limitations}

\input{related}
\input{conclusions}

{
\footnotesize

\newpage

\bibliographystyle{IEEEtranS}
\def\UrlBreaks{\do\/\do-}
\bibliography{biblio}
}

\end{document}

%% file: mymacros.tex
%
%
\newcommand{\REINSTATE}[1]{}
\newcommand{\SQUEEZE}[1]{#1}


\newcommand{\nnum}[1]{\sepnum{.}{,}{}{#1}}
\newcommand{\ttime}[1]{\sepnum{.}{,}{}{#1}}
\newcommand{\hide}[1]{}
\newcommand{\callbackExample}[2]{{#1 (#2):}}
\newcommand{\code}[1]{{\ifmmode{\mathtt{#1}}\else$\mathtt{#1}$\fi}}
\newcommand{\htmltag}[1]{\code{<\!\!#1\!\!>}}
\newcommand{\smallcode}[1]{{\footnotesize\ifmmode{\mathtt{#1}}\else$\mathtt{#1}$\fi}}
\newcommand{\verysmallcode}[1]{{\scriptsize\ifmmode{\mathtt{#1}}\else$\mathtt{#1}$\fi}}
\newcommand{\verysmallcodebf}[1]{{\scriptsize\ifmmode{\bm{\mathtt{#1}}}\else$\bm{\mathtt{#1}}$\fi}}

\makeatletter
\renewcommand{\verbatim@font}{\footnotesize\fontfamily{cmtt}\selectfont}



\newcounter{example}
\renewcommand{\theexample}{\arabic{example}}
\newenvironment{ex}{
\refstepcounter{example}{
\smallskip \noindent {\sf \\ Example \theexample:~}}}{ \ 
}

\newenvironment{verbacode}[2]{%
    \newcommand*{\mycaptiontext}{#1}%
    \newcommand*{\mylabeltext}{#2}%
    \begin{figure*}[tb]%
    \begin{centering}%
    \rule[2pt]{\linewidth}{0.6pt}
    \begin{minipage}{12cm}%
    \scriptsize%
    \begin{alltt}}{%
    \end{alltt}%
    \end{minipage}%
    \par%
    \end{centering}%
    \vspace*{2pt}%
    \rule[10pt]{\linewidth}{0.6pt}
        \vspace*{-6ex}%
    \caption{\mycaptiontext}%
    \label{\mylabeltext}%
    \end{figure*}}

\newenvironment{columnfig}[3]{%
    \newcommand*{\mycaptiontext}{#2}%
    \newcommand*{\mylabeltext}{#3}%
    \begin{figure}[#1]%
    \begin{centering}%
    \rule[2pt]{\linewidth}{0.6pt}
    \begin{minipage}{\columnwidth}}{%
    \end{minipage}%
    \par%
    \end{centering}%
    \vspace*{2pt}%
    \rule[10pt]{\linewidth}{0.6pt}
        \vspace*{-6ex}%
    \caption{\mycaptiontext}%
    \label{\mylabeltext}%
    \end{figure}}


\makeatletter

\newcommand{\zug}[1]{\langle #1 \rangle}
\newcommand \Pre    {\ensuremath{\mathit{Pre}}}
\newcommand \Post   {\ensuremath{\mathit{Post}}}
\newcommand \Prog   {\ensuremath{\mathsf{Prog}}}
\newcommand \Spec   {\ensuremath{\mathsf{Spec}}}
\newcommand \Impl   {\ensuremath{\mathsf{Impl}}}
\newcommand \Path   {\ensuremath{\mathsf{Path}}}
\newcommand \Triple  {\ensuremath{\mathsf{Triple}}}
\newcommand \PreE   {\ensuremath{\mathit{PreE}}}
\newcommand \PostE   {\ensuremath{\mathit{PostE}}}
\newcommand \wpc   {\ensuremath{\mathsf{WP}}}
\newcommand \eimpl {\ensuremath{\Rrightarrow}}
\newcommand \expify[1] {\lbrack #1 \rbrack}

\newcommand \xsrc  {\ensuremath{X_{\mathit{src}}}}
\newcommand \xsnk  {\ensuremath{X_{\mathit{src}}}}
\newcommand \xsan  {\ensuremath{X_{\mathit{src}}}}

\def\QEDclosed{\mbox{\rule[0pt]{1.3ex}{1.3ex}}}
\def\QEDopen{{\setlength{\fboxsep}{0pt}\setlength{\fboxrule}{0.2pt}\fbox{\rule[0pt]{0pt}{1.3ex}\rule[0pt]{1.3ex}{0pt}}}}
\def\QED{\QEDclosed}
\def\proof{\noindent{\itshape Proof: }}
\def\endproof{\hspace*{\fill}~\QED\\\par\endtrivlist\unskip}

\newcommand{\ab}[1]{\textbf{(AB:#1)}}

\newcommand{\point}[1]{\par\smallskip\noindent{\bf #1:}}


%% file: abstract.tex
\begin{abstract}
As malware, exploits, and cyber-attacks advance over time, so do the mitigation techniques available to the user. However, while attackers often abandon one form of exploitation in favor of a more lucrative one, mitigation techniques are rarely abandoned.
Mitigations are rarely retired or disabled since proving they have outlived their usefulness is often impossible. As a result, performance overheads, maintenance costs, and false positive rates induced by the different mitigations accumulate, culminating in an outdated, inefficient, and costly security solution.

We advocate for a new kind of tunable framework on which to base security mechanisms. This new framework enables a more reactive approach to security allowing us to optimize the deployment of security mechanisms based on the current state of attacks. Based on actual evidence of exploitation collected from the field, our framework can choose which mechanisms to enable/disable so that we can minimize the overall costs and false positive rates while maintaining a satisfactory level of security in the system.

We use real-world Snort signatures to simulate the benefits of reactively disabling signatures when no evidence of exploitation is observed and compare them to the costs of the current state of deployment. Additionally, we evaluate the responsiveness of our framework and show that in case disabling a security mechanism triggers a \emph{reappearance} of an attack we can respond in time to prevent mass exploitation.

Through a series of large-scale simulations that use integer linear and Bayesian solvers, we discover that
our responsive strategy is both computationally affordable and results in significant reductions in false positives, at the cost of introducing a moderate number of false negatives.
Through measurements performed in the context of large-scale simulations we find that the time to find the optimal sampling strategy is under~\overlapSolvedDaysTime minutes in the vast majority of cases.
The reduction in the number of false positives is significant, about~\empirical{20\%} over traces that are about~\yearsData years long ($\sim$\empirical{9.2 million} false positives).

\end{abstract}

%% file: intro.tex
\section{Introduction}
\label{sec:intro}

Much of the focus in the security community in the last several decades has been on discovering, preventing, and patching vulnerabilities. While both new vulnerability classes and new vulnerabilities are discovered seemingly every day, the exploitation landscape often remains murky. For example, despite buffer overruns, cross-site scripting~(XSS), and SQL injection attacks~(SQLIA) being heralded as the vulnerabilities of the decade~\cite{veracode-popular}, there is precious little published evidence of how commonly \emph{exploited} XSS or SQLIA might be in practice; of course, there is a number of studies~\cite{sqlia-prevalent} on how \emph{vulnerability trends} change over time. One of the studies we present in this paper suggests that, for example, XSS \emph{exploitation} is not nearly as common as would be suggested by the daily stream of discovered \emph{vulnerabilities} (found at \url{www.openbugbounty.org/} or \url{xssed.org}).

\point{Changing exploitation landscape}
The security industry produces regular reports that generally present a growing number of vulnerabilities, some in widely-deployed software. It is exceedingly tempting to misinterpret this as a growth trend in the number of actual exploits. However, the evidence for the latter is scant at best. Due to a number of defense-in-depth style measures over the last decade, including stack canaries, ASLR, XRF tokens, automatic data sanitization against XSS and a number of others, practical exploitation on a mass scale now requires an increasingly sophisticated attacker. We see this in the consolidation trends of recent years. For example, individual drive-by attacks have largely been replaced by exploit kits~\cite{kotov,pexy,kizzle16}. In practice, mass-scale attacks generally appear to be driven by a combination of two factors:~1) ease of exploitation, and~2)~whether attacks are consistently monetizable. This is clearly different from targeted attacks and APTs where the upside of a single successful exploitation attempt may be quite significant to the attacker.

Given the growing scarcity in \emph{exploitable} vulnerabilities, there is some recent evidence that attackers attempt to take advantage of publicly disclosed attacks right after their announcement, while the window of vulnerability is still open and many end-users are still unpatched; Bilge~\etal~\cite{Bilge:2012:BWK:2382196.2382284} report an increase of~\emph{5 orders of magnitude} in attack volume following public disclosures of zero-day attacks.
The situation described above leads to a long tail of attacks~---~a period of time when attacks are still possible but are increasingly rare. It is tempting to keep the detection mechanism on during the long tail. However, it is debatable whether that is a good strategy, given the downsides. We argue that the usual human behavior in light of the rapidly-changing landscape is inherently \emph{reactive}, however, often not reactive enough.

\subsection{Mounting Costs of Security Mechanisms }
One of the challenges of security mechanisms is that their various costs can easily mount if unchecked over time.
\setlength{\leftmargini}{1em}
\begin{itemize}
\itemsep=-1pt
    \item {\bf False positives.} FPs have nontrivial security implications~\cite{av-fp-sink,sukwong2011commercial}. According to a recent Ponemon Institute report~\cite{Ponemon:2015}, ``The average cost of time wasted responding to inaccurate and erroneous intelligence can average \$1.27 million annually.'' Futrthermore, `` Of all alerts, 19\% are considered
reliable \emph{but} only 4\% are investigated.''

	\item {\bf Performance overhead.}
	Our studies of scanning and filtering costs in Section~\ref{sec:av-costs} show that while the IO overhead from opening files, etc. can be high, the costs of scanning and filtering increases significantly as more signatures are added to the signature set.
When many solutions are applied simultaneously within a piece of software, the overhead of even relatively affordable mechanisms, such as stack canaries~\cite{canaries-cost} and ASLR~\cite{payer2012too}, can be \emph{additive}.
There is a growing body of evidence that security mechanisms that incur an overhead of~10\% or more do not tend to get widely deployed~\cite{war-in-memory}. However, several low-overhead solutions one on top of another can clearly
easily exceed the~10\% mark.

\hide{
	\item {\bf Infrastructure.}
	The cost of backend maintenance \emph{and} sending security updates is difficult to estimate. These may include analyzing malware, testing signatures, running honeypots, etc.~\cite{provos2009cybercrime,Provos:2008:YIP:1496711.1496712}.
	Furthermore, reducing the signature set means less data that needs to be transmitted to the end-user, thus reducing network costs and disk space requirements.
	
		\item {\bf Technical debt.}
    There is a problem of technical debt associated with \emph{maintaining} existing security mechanisms as attack volume diminishes. This is considered to be a growing problem in the machine learning community~\cite{Sculley:SE4ML:2014}, but is not often cited as an issue in security and privacy.
    For example, Kerschbaumer~\cite{firefox-csp} reports that only \emph{modernizing} Firefox's CSP (content security policy) implementation took over~125,000 lines of code over a period of~20 months by a dedicated developer. Clearly, not every project has these kinds of resources.
    When considered over a long period of time, the technical debt accumulates to a point that the software maker can no longer deal with the maintenance issues or can only do so at the expense of introducing new defense strategies, to deal with today's issues.
}

\end{itemize}

The performance and false positive costs are in addition to maintenance and update costs, which are also difficult to predict i.e. analyzing malware, testing signatures, running honeypots, etc.~\cite{provos2009cybercrime,Provos:2008:YIP:1496711.1496712}).

We argue that as a result of the factors above, over a long period of time, a situation in which we only \emph{add} security mechanisms is unsustainable due to the mounting performance costs and accumulating false positives. This is akin to performing a DOS attack against oneself; in the limit, the end-user would not be able to do much useful work because of the overhead and false positive burden of existing software.

\point{Reluctance to disable}
At the same time, actively \emph{removing} a security mechanism is tricky, if only from a PR standpoint. In fact, we are not aware of a recent major security mechanism that has been officially disabled.
One of the obvious downsides of tuning down any security mechanism is that the recall decreases as well.
However, it is important to realize that when tuning down a specific mechanism is driven by representative measurements of its effectiveness in the wild, this is a good strategy for dealing for mass attacks.

Targeted attacks, on the other hand, are likely to be able to overcome most existing defenses, as we have seen from Pwn2own competition, XSS filter bypasses~\cite{waf-bypass-bh-2016,protocol-level-evasion-2012,Bates:WWW:2010}, etc. We hypothesize that sophisticated targeted attacks are not particularly affected by existing defenses. However, reducing the level of defense may invite new waves of mass attacks, which can be mitigated by upping the level of enforcement once again. 
We therefore consider entirely taking out a security mechanism, rather than simply disabling or tuning it down, to be ill-advised.

\subsubsection{Scanning and Filtering Costs}
\label{sec:av-costs}

\begin{figure*}[h!]
\centering
\begin{subfigure}[tb]{0.3\textwidth}
\centering
\ifpdf
    \includegraphics[width=1\textwidth]{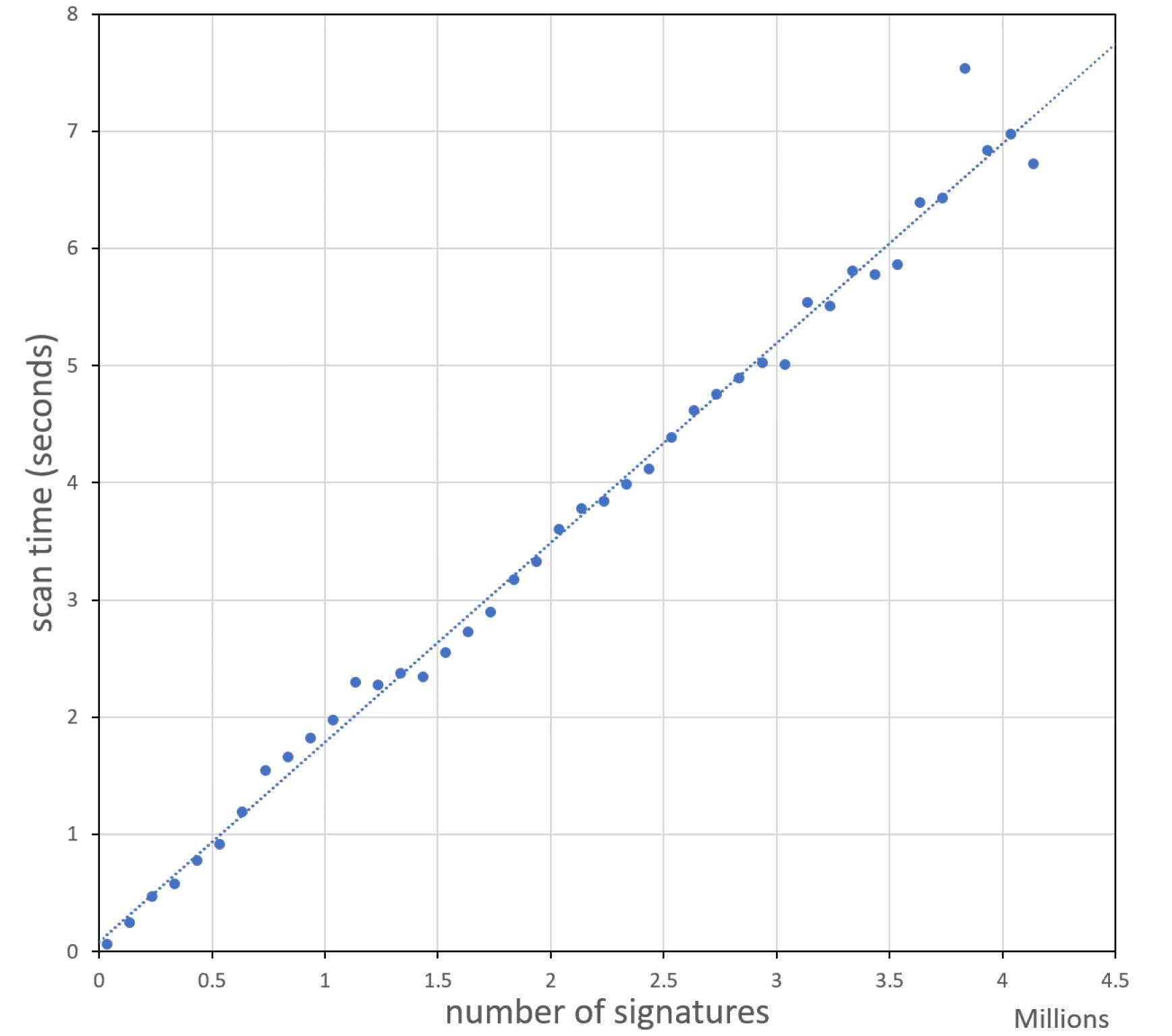}
\fi
\caption{Using ClamAV~(0.99.2) to scan a single~248~MB file.}
\label{fig:clamav}
\end{subfigure}
\begin{subfigure}[tb]{0.3\textwidth}
\centering
\ifpdf
    \includegraphics[width=1\textwidth]{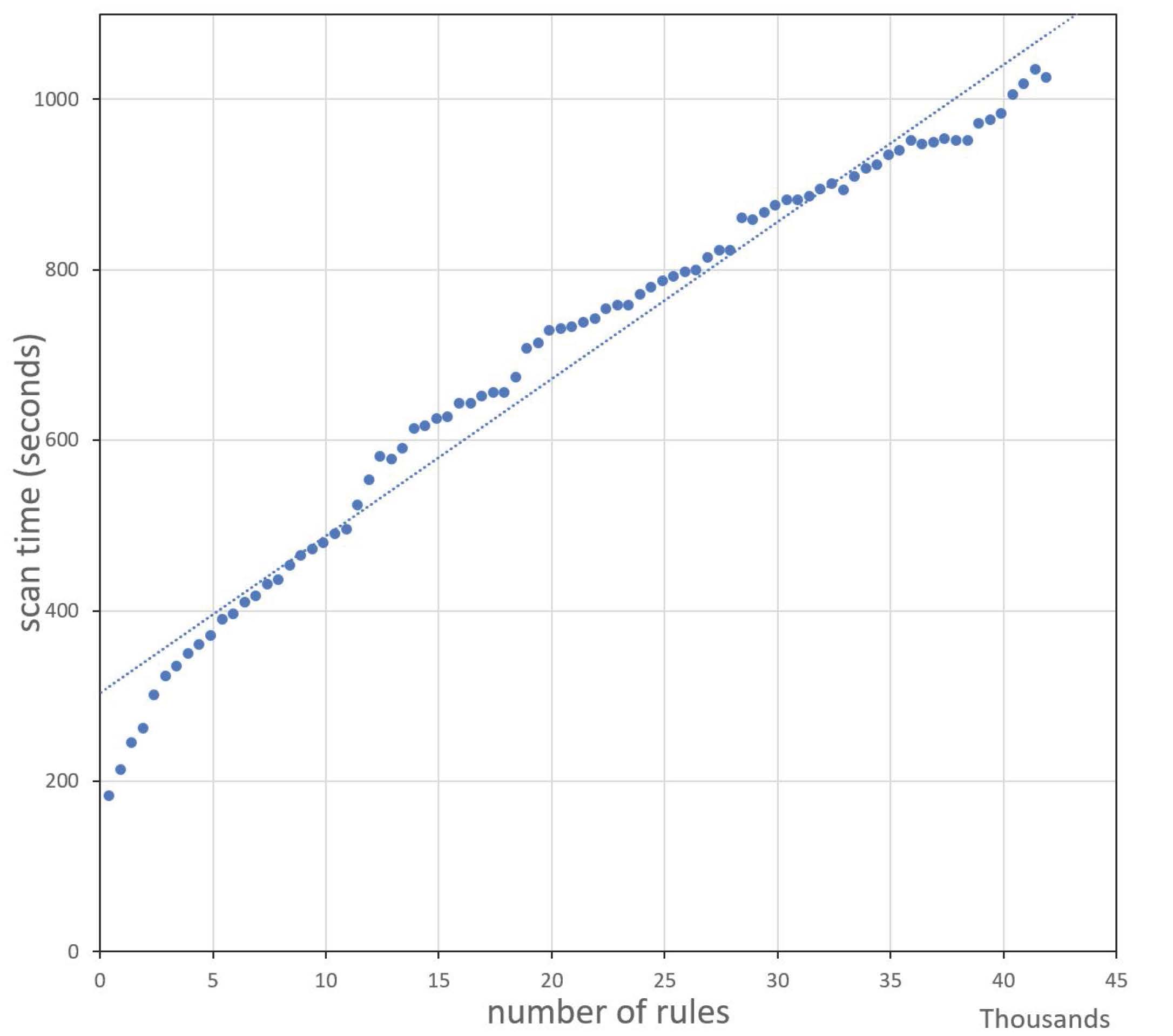}
\fi
\caption{Using Snort~(2.9.9.0) to scan a~4.1~GB pcap file.}
\label{fig:snort}
\end{subfigure}
\begin{subfigure}[tb]{0.3\textwidth}
\centering
\ifpdf
    \includegraphics[width=1\textwidth]{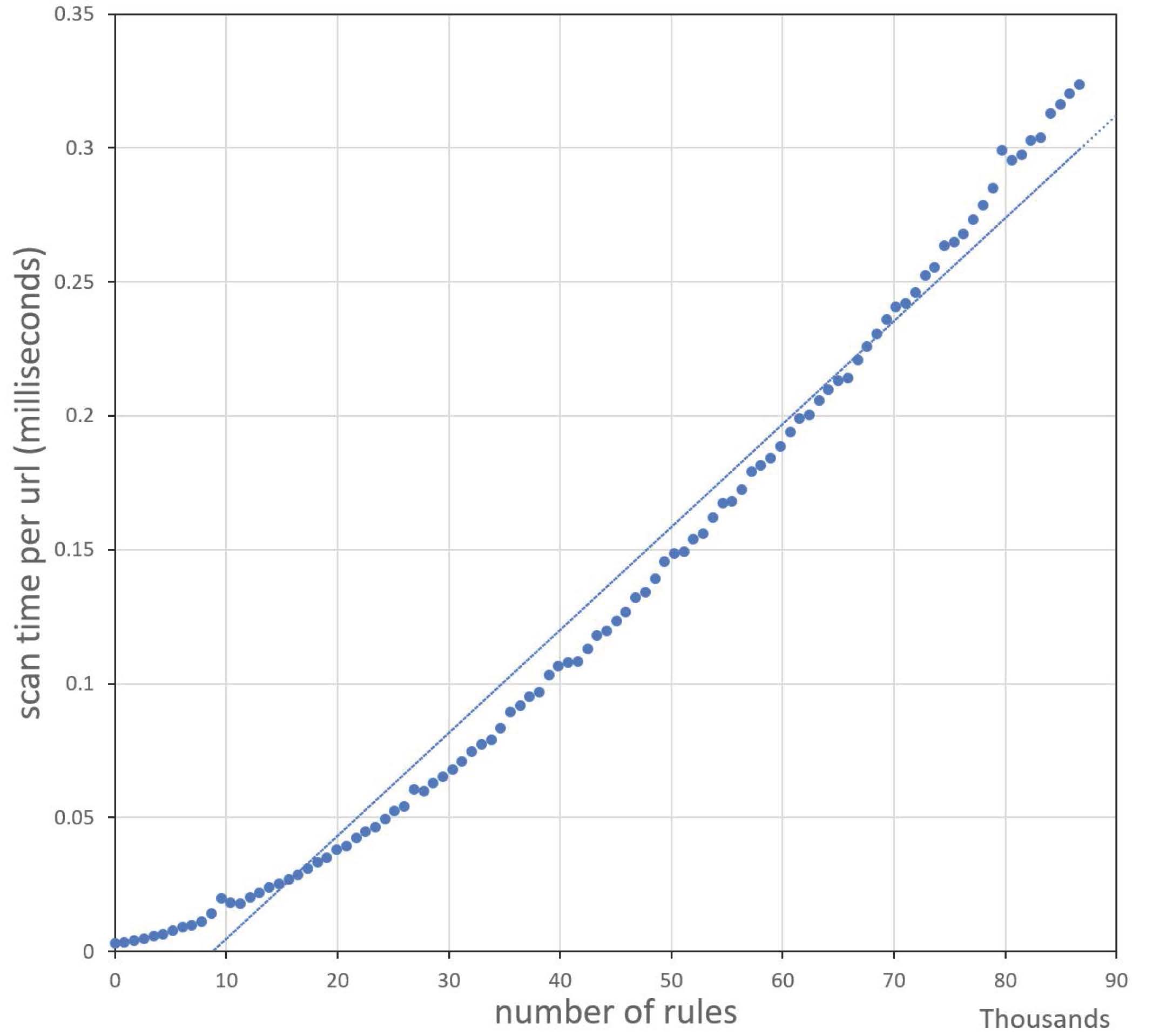}
\fi
\caption{Using Brave's adblock engine~(4.1.1) to scan 15.5k URLs.}
\label{fig:adblock}
\end{subfigure}
\caption{Average scanning time as a function of the rule/signatures set size. The dashed lines represent linear fits to the measurement.}
\end{figure*}

To demonstrate our claims of the inefficiency and mounting costs of maintaining a large amount of outdated security mechanism, we turn to 
ClamAV~\cite{clamav}, Snort~\cite{snort}, and Brave's ad blocking engine~\cite{adblock}.
\begin{itemize}
\item
We installed the latest version of the ClamAV engine~(0.99.2) and used it to scan a single file of~\empirical{248}~MB.

To understand how the running time is affected by the size of the signature set, we ran the scan several times using different sized subsets of the ClamAV signature dataset.
Figure~\ref{fig:clamav} shows the obtained scan times, in seconds.

\item
We performed similar measurements using the Snort NIDS engine. 
Using the latest version of Snort~(2.9.9.0), we scanned a collection of pcap files,  which we obtained from~\cite{pcap}, with signature sets of different sizes.
Figure~\ref{fig:snort} shows the results of these measurements.

\item Lastly, we used the ad-block engine of the Brave browser~\cite{brave, adblock} to the effect the ad-block rule set size has on the average scan time per url.
We used rules from the EasyList and EasyPrivacy rule sets~\cite{easylist}, totaling at~ 86,665 rules.
Utilizing the Brave ad-block engine~(4.1.1), we scanned a set of URLs containing~15,507 elements. We ran the experiment using different-sized subsets of the rule set.
Figure~\ref{fig:adblock} shows the obtained average scan times per URL, in milliseconds.
\end{itemize}
All three of the above figures clearly exhibits an approximately \emph{linear} correlation between the rule/signature set size and the average scan time.
These charts demonstrate the hidden cost over time of adding signatures and rarely removing them. This factor together with false positives argues for removing signatures more aggressively.

\subsection{Toward Tunable Security}
In recent years we have seen growing evidence that vulnerability statistics and exploit statistics are at odds.
Nayak~\etal~\cite{exploited-in-wild-raid} report that despite an increase in reported vulnerabilities in the last several years,  while the amount of exploitation goes down. Furthermore, only a portion of vulnerabilities (about~35\%) actually gets exploited in a practical sense. Furthermore, vulnerability severity rankings are often misleading, as they do not necessarily correlate with the attractiveness or practicality of exploitation~\cite{misleading-severity}. 
Barth~\etal~\cite{Barth:FC:2010} advocate a reactive approach to security where previous attack patterns are used to decide how to spend the defender's finite budget.

\point{Methodology}
In this paper, we experimentally prove the advantages of reactive security. We look at widely deployed security mechanisms, where the potential, for example, of a single false positive is amplified by the potentially vast installation base. A good example of such a system is an anti-virus (AV) engine or an IDS. The reactive approach to security is also supported by the number of zero-days that are observed in the wild and reported by Bilge~\etal~\cite{Bilge:2012:BWK:2382196.2382284}.
Our experimental evaluation in Section~\ref{sec:expt} is based on~\yearsData years of Snort signatures. We argue that a well-calibrated simulation is the best way we have to assess the reactive security mechanism proposed in this paper.
We use an extremely valuable study of over~75 million real-world attacks collected by Symantec~\cite{Allodi2015} from~2015 to calibrate our simulation.

\point{Tunable security mechanism}
We propose an alternative: a tunable mechanism, guided by a strategy that allows the defender to selectively apply the mechanism based on internal and/or external sets of conditions.
For example, a centralized server can notify the client that certain categories of attacks are no longer in the wild, causing the client to reduce their sampling rates when it comes to suspicious traffic. Signatures are thus retired when the threats they are designed to look for become less prevalent. For this to work, there needs to be a greater level of independence between the security mechanism and the policy, the kind of separation that is already considered a major design principle~\cite{mechanism-policy-separation,policy-hydra}.

\point{Beyond signatures}
It important to emphasize that while we primarily experiment with signatures in this paper, the ideas apply well beyond signatures. Specifically, most runtime security enforcement mechanisms, albeit not all, can be turned off, either partially or entirely. Mechanisms that rely on matching lists for their enforcement include XSS filters~\cite{Bates:WWW:2010} and ad blockers~\cite{Berlin2015,Merzdovnik,Mughees}, to name just a few. Similarly, one can apply XFI or CFI to some DLLs, but not others, that have not been implicated in recent attacks. Of course, the ability to adjust which DLLs are CFI-protected depends greatly on having a dynamic software deployment infrastructure, which we claim to be a desirable goal.

\point{Matching today's reality}
In many ways, our proposal is aligned with practical security enforcement practices of adjusting the sensitivity levels for detectors depending on the FP-averseness of the environment in which the detector is deployed. The Insight reputation-based detector from Symantec allows the user to do just that~\cite{symantec-reputation}.
Our ultimate goal here, of course, is to reduce the factors listed above, i.e. the false positive rate and the performance overhead.

\subsection{Contributions}
Our paper makes the following contributions:
\begin{itemize}\itemsep=-1pt
    \item\textbf{\Tunable.}
    We point out that today's approach to proactive security leads to inflexible and bloated solutions over time. We instead advocate a novel notion of \tunable security design, which allows flexible and fine-grained policy adjustments on top of existing security enforcement mechanisms. This way, the protection level is tuned to match the current threat landscape and not either the worst-case scenario or what that landscape might have been in the past.
    \item\textbf{Sampling-based approach.} For a collection of mechanisms that can be turned on or off independently, we propose a strategy for choosing which mechanisms to enable for an optimal combination of true positives, false positives, and performance overheads.
    \item\textbf{Formalization.} We formalize the problem of optimal adjustment for a mechanism that includes an ensemble of classifiers, which, by adjusting the sampling rates, produces the optimal combination of true positives, false positives, and performance overheads.
    \item\textbf{Simulation.} Using a simulation based on a history of Snort signature updates over a period of about~\yearsData years, we show that we can adjust the sampling rates within a window of minutes. This means that we can rapidly react to a fast-changing exploit landscape.
\end{itemize}
\hide{
This is a paper about better ways to build security mechanisms and to more effectively tune security policies. We have chosen to evaluate it by doing similaiton on historical data. We realize that this approach is imperfect: it would be more convincing to deploy our flexible tinable technique, however, the only people who can really do so are those who deploy protection mechanisms daily, as the threat landscape changes, such as AV or malware protection companies. Given that we are not employed in such a context and the difficulty of deploying these mechanisms otherwise, we feel we are doing our best with the help of historical data to make our argument.
} 

\hide{
\subsection{Paper Organization}
The rest of the paper is organized as follows.
Section~\ref{sec:background} gives an overview of the exploitation landscape.
Section~\ref{sec:sampling} defines an optimization problem that improves the true positive rates and reduces the false positive rate and enforcement costs, while favoring higher-severity warnings.
Section~\ref{sec:expt} describes our experimental evaluation.
Section~\ref{sec:limitations} talks about the limiations of our approach. 
Finally, Sections~\ref{sec:related} and~\ref{sec:conc} describe related work and conclude. 
}

%% file: background.tex
\newcommand{\et}{{\textsc{Emerging Threats}}\xspace}

\section{Background}
\label{sec:background}

When it comes to malware detection, anti-virus software has long been the first line of defense. However, for almost as long as AV engines have been around, they have been recognized to be far from perfect, in terms of their false negative rates~\cite{tenable, gashi2009experimental}, false positive rates, and, lastly, in terms of performance~\cite{real-time-av,Dien2014}.

Clearly, the choice of signature database plays a decisive role in the success of the AV solution. This is illustrated by the promises of SaneSecurity (\url{http://sanesecurity.com}) to deliver improved detection rates, by using the free open-source ClamAV detection engine and their own carefully curated and frequently updated database of signatures. For example, as of August~2016, they claim a detection rate of~97.11\% vs only~13.82\% for out-of-the-box ClamAV, using a database of little over~4,000 signatures vs. almost~250,000 for ClamAV.

\hide{
Several researchers have proposed generating new signatures automatically~\cite{newsome2005polygraph,perdisci2010behavioral,griffin2009automatic,sathyanarayan2008signature,zolkipli2010framework}. Signature \emph{addition} seems to largely remain a manual process, supplemented with testing potential AV signatures against known deployments, often within virtual machines.
}

While the issue of false negatives is generally known to industry insiders, the issue of false \emph{positives} receives much more negative press. Reluctance to remove or disable older signatures runs the risk of unnecessarily triggering false positives, which are supremely costly to the AV vendor, as reported by AV-Comparatives~\cite{av-fp-sink} (\url{http://www.av-comparatives.org}).

\subsection{The Changing Attack Landscape}
When the attacks for which some mitigation mechanism was designed are no longer observed in the wild, it might seem very alluring to remove said mechanism.
However, in most real world scenarios, we are not able to fully retire mitigation techniques.
Before disabling some mitigation mechanism, one should examine the reason these attacks are no longer being observed and whether this is simply due to the observational mechanism and data collection approach being faulty.

Today, large-scale exploitation is often run as a business, meaning it is driven largely, but not entirely, by economic forces.
The lack of observed attacks might simply be associated with an increased difficulty in monetizing the attack.
Given that the attacker is aiming to profit from the attack, if the cost of mounting a successful attack is too high compared to either the possible gains or alternative attack vectors, the attacker will most likely opt not to execute it as it is no longer cost-effective. We see these forces in practice as the attack landscape changes, with newer attacks such as ransomware becoming increasingly popular in the last several years while older attacks leading to the theft of account credentials becoming less common because of two-factor authentication, geo-locating the user, etc.
To summarize, we identify two common cases where monetizing becomes hard.
\begin{itemize}\itemsep=-1pt
	\item
\textbf{No longer profitable.} The first is the result of causes other than the mitigation mechanism.
For example, when clients are no longer interested in the possible product of the attack or if there are other security mechanisms in place that prohibit the usefulness of the attack's outcome.
In such cases, removing the mitigation mechanism in question will most likely not have a practical negative effect on the system since the attack remains not cost-effective.
    \item
\textbf{Effective mitigation.} The other case is when the attack is not cost-effective due to difficulties imposed by the mitigation mechanism.
In such cases, removing the mechanism will result in an increase in the cost-effectiveness of the attacks it was aimed to prevent.
This might result in the reemergence of such attacks.
\end{itemize}
By sampling the relative frequency of attacks of a particular kind, we cannot always determine which case we are currently faced with. It may also be a combination of these two factors. However, there is growing evidence that attackers are frequently going after the low-hanging fruit, effectively behaving lazily~\cite{LazyAttacker}.

We therefore suggest an alternative that acts as a middle ground by introducing sampling rates for all mitigation mechanisms.
In the first case, the mitigation mechanism is no longer needed, therefore adding a sampling rate will not reduce the security of the system, but will provide fewer benefits than a complete removal.
On the other hand, in the second case, the security of the system is somewhat lowered, but the statistical nature of the sampling rate maintains some deterrence against attackers.

\subsection{Study of Snort Signatures}
\begin{figure*}[h!]
\centering
\begin{subfigure}[tb]{1\textwidth}
\centering
\ifpdf
    \includegraphics[width=\textwidth]{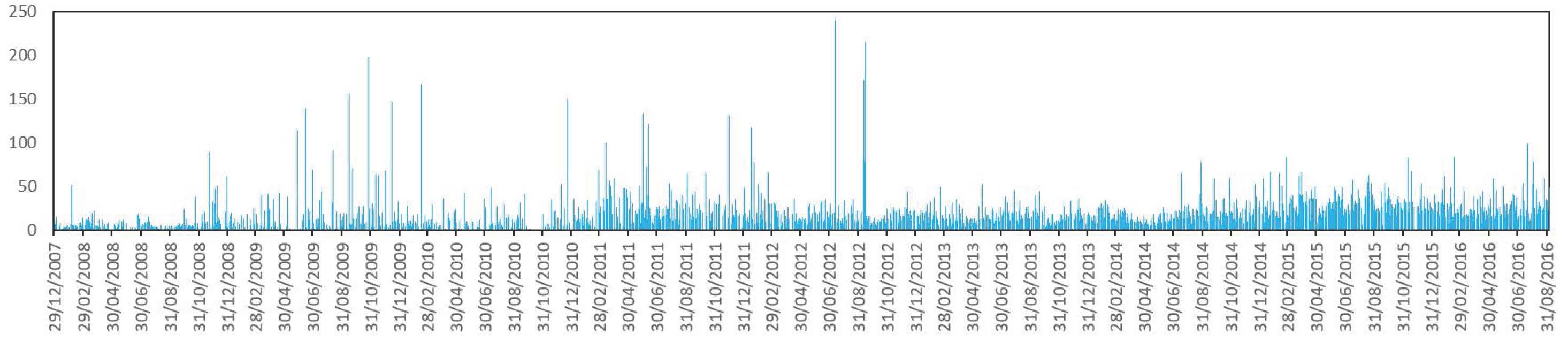}
\fi
\caption{Additions of signatures in the Snort emerging threats database.}
\end{subfigure}
\\

\begin{subfigure}[tb]{1\textwidth}
\centering
\ifpdf
    \includegraphics[width=\textwidth]{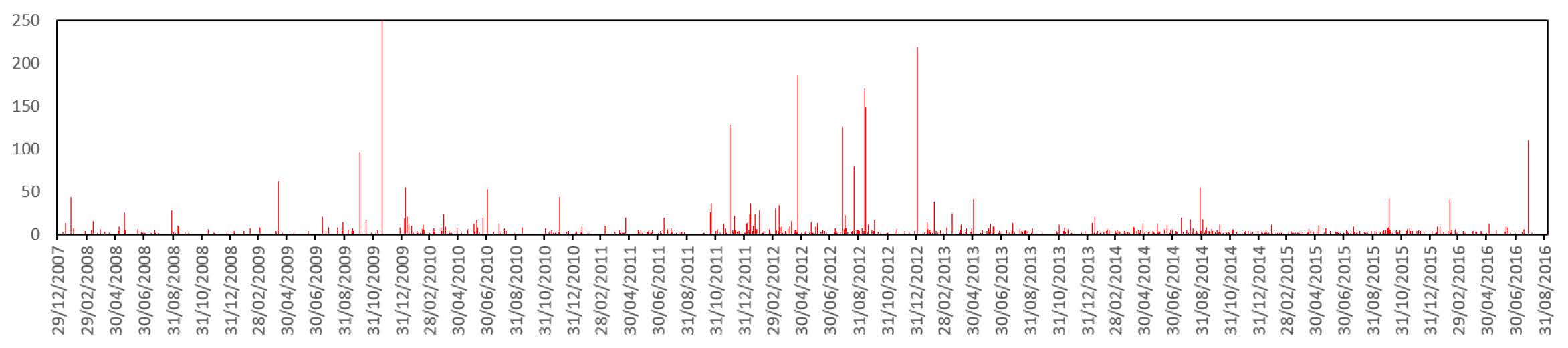}
\fi
\caption{Removals of signatures in the Snort emerging threats database.}
\end{subfigure}\\

\begin{subfigure}[tb]{1\textwidth}
\centering
\ifpdf
    \includegraphics[width=\textwidth]{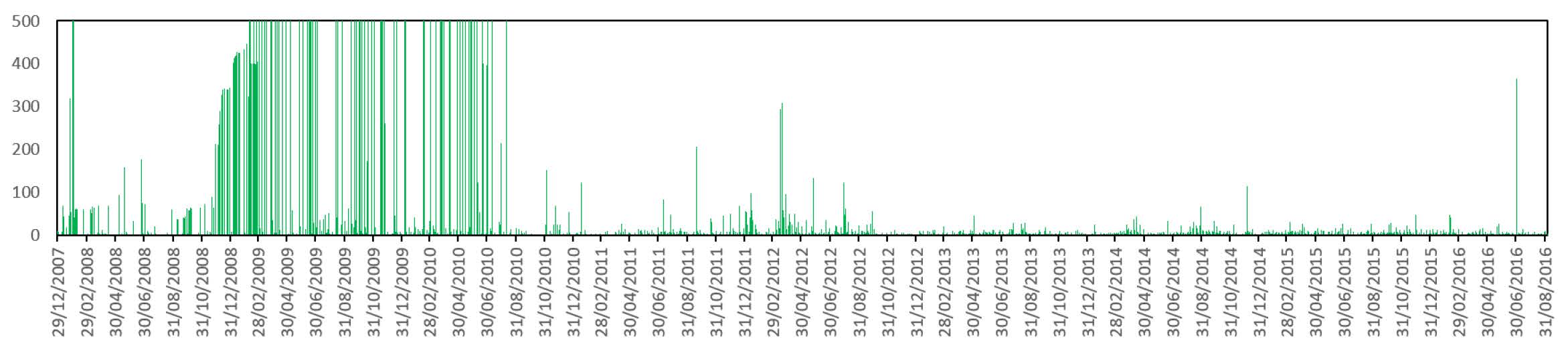}
\fi
\caption{Updates of signatures in the Snort emerging threats database.}
\end{subfigure}

\caption{Dynamics of Snort signatures between~12/30/2007 and~9/6/2016.}
\label{fig:snort-add-remove}
\end{figure*}

To test some of these educated guesses, we have performed an in-depth study of Snort signatures. Focusing on the dynamics of signature addition and removal, we have mined the database of Snort signatures, starting on~12/30/2007 and ending on~9/6/2016. Daily updates to the Snort signature database are distributed through the \et mailing list archived at \url{https://lists.emergingthreats.net}, which we used to determine which signatures, if any were 1) added, 2) removed, or 3) modified every single day. Figure~\ref{fig:snort-add-remove} presents statistics of Snort signatures obtained by exploring the dataset we collected by crawling the mailing list archive.

It is evident from the figure that there are many more signature additions than removals. These statistics support our claim that signature dataset sizes are growing out of control and becoming unsustainable.

\begin{figure}[tb]
\centering
\tiny
\begin{Verbatim}
2023020 - ET TROJAN ProjectSauron Remsec DNS Lookup (rapidcomments.com) (trojan.rules)
2023021 - ET TROJAN ProjectSauron Remsec DNS Lookup (bikessport.com) (trojan.rules)
2023022 - ET TROJAN ProjectSauron Remsec DNS Lookup (myhomemusic.com) (trojan.rules)
2023023 - ET TROJAN ProjectSauron Remsec DNS Lookup (flowershop22.110mb.com) (trojan.rules)
2023024 - ET TROJAN ProjectSauron Remsec DNS Lookup(wildhorses.awardspace.info) (trojan.rules)
2023025 - ET TROJAN ProjectSauron Remsec DNS Lookup (asrgd-uz.weedns.com) (trojan.rules)
2023026 - ET TROJAN ProjectSauron Remsec DNS Lookup (sx4-ws42.yi.org)(trojan.rules)
2023027 - ET TROJAN ProjectSauron Remsec DNS Lookup (we.q.tcow.eu)(trojan.rules)
\end{Verbatim}
\caption{Connecting signatures to ProjectSauron malware (\code{http://bit.ly/2eX1O9h}).}
\label{fig:sauron-malware}
\hide{
\end{subfigure}
\\[2mm]
\begin{subfigure}{\columnwidth}
\tiny
\begin{Verbatim}
2003182 || ET TROJAN Prg Trojan v0.1-v0.3 Data Upload || URL
2003183 || ET TROJAN Prg Trojan Server Reply || URL
2003184 || ET TROJAN Prg Trojan v0.1 Binary In Transit || URL
2003185 || ET TROJAN Prg Trojan v0.2 Binary In Transit || URL
2003186 || ET TROJAN Prg Trojan v0.3 Binary In Transit || URL
2007688 || ET TROJAN Prg Trojan HTTP POST v1 || URL
2007724 || ET TROJAN Prg Trojan HTTP POST version 2 || URL
    URL = url,www.securescience.net/FILES/securescience/10378/pubMalwareCaseStudy.pdf
\end{Verbatim}
\caption{Zeus (Prg) malware (\code{http://bit.ly/2bIS3hk}).}
\label{fig:zeus-malware}
\end{subfigure}
\caption{Connecting signatures to known malware }
}
\squeezeup
\end{figure}

Below we give several representative examples of signature addition, update, and removal.
\begin{ex}{\bf Signature addition.}
Figure~\ref{fig:sauron-malware} shows the addition of new signatures in response to observations of the ProjectSauron~\cite{project-sauron} malware in the wild. The connection between the malware and \et signatures designed to prevent it is evident from the signature description.
%
\end{ex}

\begin{ex}{\bf Signature update.}
\hide{
Figure~\ref{fig:dhl-update} shows an example of a typical mailing list exchange leading up to a signature change.
\begin{figure}[tb]
\tiny
\begin{Verbatim}
> We have 2010148 already:
> content:"Content-Disposition|3A| attachment|3b|"; nocase;
> content:"filename"; within:100; content:"DHL_"; nocase;
> within:50;
> pcre:"/filename\s*=\s*\"DHL_(Label_|document_|
> package_label_|print_label_).{5,7}\.zip/mi";
>
> I'll modify to fit the new style. The old is gone!
>
> Thanks Jason!
>
> Matt
>
> On Nov 3, 2010, at 10:35 AM, Weir, Jason wrote:
>
> > Seeing these as inbound smtp attachments
> >
> > DHL_label_id.Nr21964.zip
> > DHL_label_id.Nr48305.zip
> > DHL_label_id.Nr3139.zip
> > DHL_label_id.Nr15544.zip
> > DHL_label_id.Nr7085.zip
> >
> > How about this for current events
> >
> > alert tcp $EXTERNAL_NET any -> $SMTP_SERVERS 25 (msg:"ET
> > CURRENT_EVENTS DHL Spam Inbound"; flow:established,to_server;
> > content:"Content-Disposition|3a| attachment|3b|"; nocase;
> > content:"filename=|22|DHL_label_id."; nocase;
> > pcre:"/filename=\x22DHL_label_id\.Nr[0-9]{4,5}\.ZIP\x22/i";
> > classtype:trojan-activity; sid:xxxxxxx; rev:0;)
\end{Verbatim}
\caption{Updating a signature based on customer feedback.}
\label{fig:dhl-update}
\end{figure}
%
}
In the case of the signature marked as 2011124, we see a false positive report about traffic on port 110 on 04/04/2016, which receives a response from the maintainers within two days:
{
\tiny
\begin{Verbatim}
we get quite a lot of false positives with this one due to
the POP3 protocol on port 110, it would be great if port 110
or more generally POP3 traffic could be excluded from this rule
-- JohnNaggets - 2016-04-02
Thanks, we'll get this out today!
-- DarienH - 2016-04-04
\end{Verbatim}
}
\noindent The maintainers added port~110 the same day they responded, resulting in this signature revision~19:
{
\tiny
\begin{Verbatim}
alert ftp $HOME_NET ![21,25,110,119,139,445,465,475,587,902,1433,2525] ->
    any any (msg:"ET MALWARE Suspicious FTP 220 Banner on Local Port (spaced)";
    flow:from_server,established,only_stream; content:"220 ";
    depth:4; content:!"SMTP"; within:20;
    reference:url,doc.emergingthreats.net/2011124;
    classtype:non-standard-protocol; sid:2011124; rev:19;)$
\end{Verbatim}
}
\end{ex}

\begin{figure}[tb]
\centering
\ifpdf
    \includegraphics[width=.9\columnwidth]{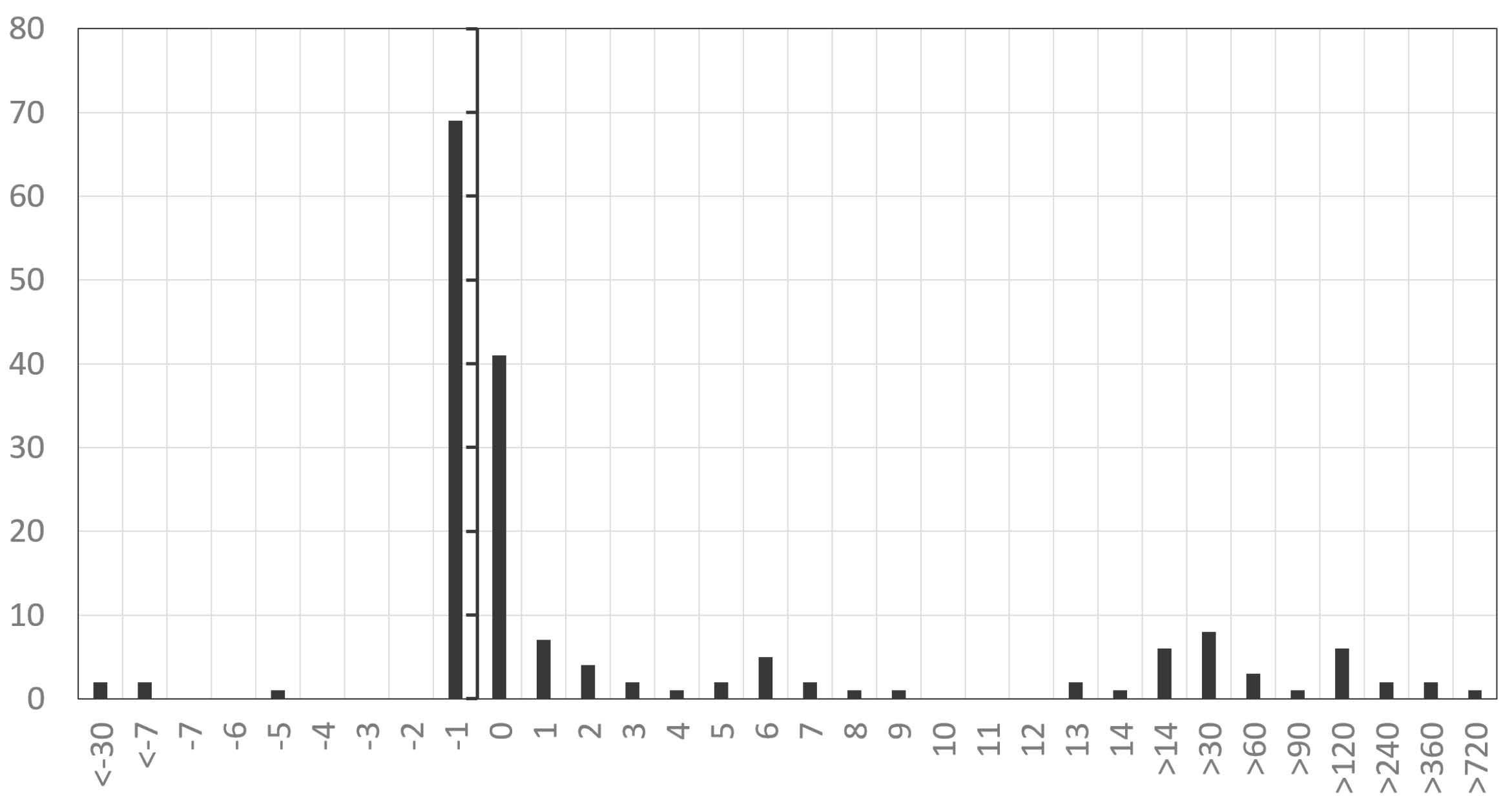}
\fi
\caption{Number of days between CVE announcement and corresponding signature addition.}
\label{fig:delay}
\squeezeup
\end{figure}

\hide{
\begin{verbatim}
     -> Added to emerging-exploit.rules (5):
        #by Rich Rumble
        #PWDump6
        #FGDump
        #This should catch both FGDump and PWDump
        #by Chandan at Secpod.com
\end{verbatim}
}

\point{CVE Additions: Evaluation of the Delay}
Another question to consider is how fast signatures for known threats are added after the threats are discovered or publicly announced. To estimate that, we have correlated~\numTotalCVEs CVEs to~\numTotalSignatures~\et signatures. This process usually involves analyzing the comments embedded in the signature to find CVE references (for example, \texttt{reference:cve,2003-0533}). We have plotted a histogram (Figure~\ref{fig:delay}) that shows the \emph{delay} in days between the CVE (according to the NIST NVD database) and the signature introduction date. As we can see, many signatures are created and added the same day the CVE is disclosed. Sadly, in quite a significant percentage of cases, signatures are added two weeks and more \emph{after} the CVE release date. What is perhaps most surprising is that many signatures are created quite a bit \emph{before} the disclosure, in many cases the day before, and in some cases over a month prior to it. This is due to other information sources that lead to signature generation.


%
%
%

\hide{
\subsection{Scanning costs}
It is important to note that scanning costs clearly increase as we add more signatures.
Appendix~\ref{sec:av-costs} describes experiments with ClamAV and Snort that show that this increase is linear.
Reducing the amount of scanned signatures thus directly improves scan performance.
}

%% file: sampling.tex
\section{Optimally Setting Sampling Rates}
\label{sec:sampling}

In this section we set the stage for a general approach to selecting sampling rates in response to changes to the data. We evaluate these ideas with practical multi-year traces in Section~\ref{sec:expt}.
We start with a model in which we have a set of classifiers (in the context of Snort signatures discussed in Section~\ref{sec:background}, each signature is a classifier) at our disposal and we need to assign a sampling rate to each of them, so as to match our optimization goals. These goals include higher true positive rates, lower false positive rates, and lower overheads\hide{, and a higher level of severity of the detected threats}. We formalize this as problem of selecting a bit-vector $\alpha,$ indicating the sampling rate for each classifier.

\subsection{Active Classifier Set}
\label{formalizing}

We assume that our classifiers send some portion of the samples they flagged as malicious for further analysis. This matches how security products, such as those from Symantec, use user machines for in-field monitoring of emerging threats.
In the context of a large-scale deployment, this results in a large, frequently updated dataset, which consists solely of true positive and false positive samples.
We can use this dataset to evaluate the average true positive, false positive, true negative, and false negative rates, induced by sampling rates for our classifiers.
Specifically, our aim is to choose a sampling bit-vector $\bar{\alpha}$ that will keep the true positive and true negatives above some threshold, while keeping the false positive rate, false negatives, and performance costs below some maximum acceptable values. We found this formulation to be most useful in our evaluation.

\point{Constraints}
Formally, these goals can be specified as a set of inequalities, one  for each threshold:
\begin{subequations}\label{goals}
    \begin{equation}\label{tp_goal}
    TP(\bar{\alpha}) \ge X_p
    \end{equation}
    \begin{equation}\label{tn_goal}
    TN(\bar{\alpha}) \ge X_n
    \end{equation}
    \begin{equation}\label{fp_goal}
    FP(\bar{\alpha}) \le Y_p
    \end{equation}
    \begin{equation}\label{fn_goal}
    FN(\bar{\alpha}) \le Y_n
    \end{equation}
    \begin{equation}\label{cost_goal}
    Cost(\bar{\alpha}) \le Z
    \end{equation}
\end{subequations}

\point{Parametrization}
Given a dataset $D$ and a set of classifiers $C$ we define the following parametrization:
\begin{itemize}\itemsep=-1pt
\item $D_i$ is the $i^{th}$ entry in the dataset and $C_j$ is the $j^{th}$ classifier;
\item $G \in (0/1)^{|D|}$, such that $G_i$ is 1 iff $D_i$ is a malicious entry in the ground-truth;
\item $R \in (0/1)^{|D|\times|C|}$, such that $R_{i,j}$ is 1 if $D_i$ is classified as malicious by $C_j$ or 0 otherwise;
\item $P \in \mathbb{R}^{|C|}$, such that $P_j$ is the average cost of classifying an entry from the dataset using $C_j$;
\item $\alpha \in [0,1]^{|C|}$, such that $\alpha_j$ is the sampling rate for classifier $c_j$.
\end{itemize}
For each set sampling rate $\alpha$ we can compute the average cost of executing the entire set of classifiers on an entry from the dataset as:
\begin{equation}
Cost(\alpha) = P^T \cdot \bar{\alpha}
\end{equation}

\point{Optimization}
To evaluate the true positive and false positive rates induced by a sampling rate $\alpha,$ we first need to evaluate the probability that an entry will be classified as malicious.
Given a constant $R$, this probability can be expressed as:
\begin{equation}\label{probability}
Pr_i(\alpha) = 1-\Pi_{j=0}^{|C|}(1-R_{i,j}\cdot\alpha_j)
\end{equation}
Based on this probability, we can express the true/false-positive rates as:
\begin{subequations}\label{noweights}
\begin{equation}\label{tp}
TP(\bar{\alpha}) = \frac{\Sigma_{i=0}^{|D|}(G_i \cdot Pr_i(\alpha))}{\Sigma_{i=1}^{|D|}(G_i)}
\end{equation}
\begin{equation}\label{fp}
FP(\bar{\alpha}) = \frac{\Sigma_{i=0}^{|D|}((1-G_i) \cdot Pr_i(\alpha))}{\Sigma_{i=1}^{|D|}(G_i)}
\end{equation}
\begin{equation}\label{tn}
TN(\bar{\alpha}) = \frac{\Sigma_{i=0}^{|D|}((1-G_i) \cdot (1-Pr_i(\alpha)))}{\Sigma_{i=1}^{|D|}(1-G_i)}
\end{equation}
\begin{equation}\label{fn}
FN(\bar{\alpha}) = \frac{\Sigma_{i=0}^{|D|}((G_i \cdot (1-Pr_i(\alpha)))}{\Sigma_{i=1}^{|D|}(1-G_i)}
\end{equation}
\end{subequations}
%
In practice, not all suggested goals are always necessary and not all goals are always meaningful.
Finding the optimal sampling rate usually depends on the setting for which it is needed.
\hide{
We note that, because $0 \le \alpha_j \le 1$ for all components of~$0 \le j < |C|$, $TP(\bar{\alpha})$ and $FP(\bar{\alpha})$ are both monotonically increasing (and so is $Cost(\bar{\alpha})$) while $TN(\bar{\alpha})$ and $FN(\bar{\alpha})$ are monotonically decreasing.
Based on this observation, each of our goals from equation~\ref{goals} determines a subspace of $[0,1]^{|C|}$.
The intersection of these subspaces results in a convex subspace of valid assignments to $\bar{\alpha}$ that satisfies all goals.
}
Next we discuss a few hypothetical scenarios and which approaches might best suit them.

\point{Prioritized objectives}
When the user can state that one objective is more important than others, a multi-leveled optimization goal can be used. In such a solution, the objective with the highest priority is optimized first. If there is more than a single possible solution, the second objective is used to choose between them, and so on.

We note that in our scenario it is extremely unlikely that one can reduce the sampling rate of a classifier without affecting the true-positive and false-positive rates. As a result, using strict objectives, such as maximize true-positives, would result in a single solution, often enabling all classifiers completely (or disabling all, depending the chosen objective). Therefore it is recommended to phrase the objectives as ``maintain X\% of true-positives'', so that some flexibility remains.

\point{Budget-aware objectives}
Often when assessing the effect a security mechanism has on a company's budget, a cost is assigned to each false positive and each false negative produced by the mechanism.
These assessments can be used to minimize the total budgetary effect of the mechanism and expected expenses.
Assuming $Cost_{FN}$ and $Cost_{FP}$ are the costs of false negatives and false positives respectively, we can express the expected expenses as:
\begin{equation}
Expenses(\alpha) = Cost_{FN} \cdot FN(\alpha) + Cost_{FP} \cdot FP(\alpha)
\end{equation}
Using this formulation we can:
\begin{itemize}\itemsep=-1pt
\item Define a budget, $Expenses(\alpha)\le BUDGET$, as a strict requirement from any sampling rate.
\item Define our problem as a standard optimization problem with the objective $minimize Expenses(\alpha)$.
\end{itemize}

\point{Balancing true positives and false positives}
In this scenario, we correlate our sampling rate optimization problem to a standard classifier optimization setting, such that our true-positive rate is equivalent to the classifier's precision while the false-positive rate becomes the recall.

In such cases, a ROC curve induced by different sampling rates can be used to select the best rate.
Taking some inspiration from the well-known $F1$-score, a similar score, $F1_{sr}$, expressed in formula~\ref{f1score} can be used to transform our problem to a single-objective optimization problem.
\begin{equation}\label{f1score}
F1_{sr}=2\cdot\frac{TP\cdot FP}{TP+FP}
\end{equation}

For efficiency, we split the process of classifier sampling rate optimization into two steps.
Real-world data often contains classifier overlap, that is, samples that are flagged by more than one classifier.
We split our dataset into batches based on the classifier overlap, so that the samples in each batch are flagged by the same set of classifiers.
Each batch is associated with true positive and false positive counts.
The first step consists of choosing the batches that are cost-effective.

\subsection{0/1 Sampling}
\label{chooseSamples}

Based on the desired optimization objective and the estimated cost ratio between false negatives and false positives (if applicable), we proceed to define the problem of finding the optimal subset of sample batches as a \emph{linear programming} problem.
At this stage, since each batch is determined by a specific classifier overlap, there is no overlap between the batches.
Therefore, the computation of the true/false positives/negatives becomes a simple summation of the associated true/false positive counts.
For example, the total true positive count is the sum of true positives associated with batches that are determined as enabled (meaning they should be sampled) and the total false negative count is the sum of the true positive counts of disabled batches.

To encode this problem, we assign each batch $b_i$ with a boolean variable $v_i$, representing whether or not the batch should be active.
We then encode the optimization goal using these variables and the associated counts.
We use a linear programming solver called Pulp~\cite{pulp}, which finds an assignment to \mbox{$\bar{v}=\{v_1,v_2,...\}$} that optimizes the optimization objective.
The output of this step is a division of the samples into enabled (meaning the classifier should sample them) and disabled samples.

When the dataset contains \emph{no classifier overlap}, meaning each sample is sampled by exactly one classifier, the output of the first step can be used as the classifier sampling rates.
In this setting, the batches essentially correspond to single classifiers and therefore disabled batches correspond to classifiers that are deemed not cost-effective according to the optimization objective.
The optimal solution in this case would be to fully enable all cost-effective classifiers and fully disable the rest.

\subsection{Inferring Sampling Rates}

In the second step of our solution, given a set of enabled samples and a set of disable samples as described in Section~\ref{chooseSamples}, we infer sampling rates for all classifiers that will induce the desired separation.

The key insight we use to infer the classifier sampling rates is to express our problem in the form of \emph{factor graphs}~\cite{factor-graphs}.
Factor graphs are probabilistic graphical models composed of two kinds of nodes: variables and factors.
A variable can be either an evidence/observation variable, meaning it's value is set, or a query variable, whose value needs to be inferred.
A factor is a special node that defines the relationships between variables. For example, given variable $A$ and $B$, a possible factor connecting them could be $A\rightarrow{B}$.

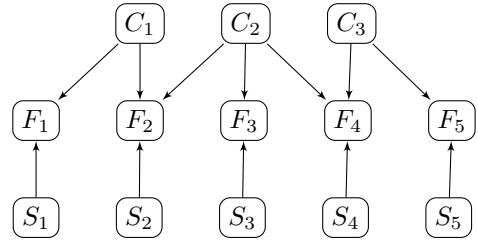
\begin{figure}
\centering
\begin{tikzpicture}[decision/.style={draw=none,fill=none},
                    block/.style   ={rectangle, draw, text centered, rounded corners},
                    line/.style    ={draw,-latex'},
                    node distance = 0.75cm,
                    auto]
    \node [block]  (s1) {$S_1$};
    \node [block, right= of s1] (s2) {$S_2$};
    \node [block, right= of s2] (s3) {$S_3$};
    \node [block, right= of s3] (s4) {$S_4$};
    \node [block, right= of s4] (s5) {$S_5$};
    \node [block, above= of s1]  (f1) {$F_1$};
    \node [block, right= of f1] (f2) {$F_2$};
    \node [block, right= of f2] (f3) {$F_3$};
    \node [block, right= of f3] (f4) {$F_4$};
    \node [block, right= of f4] (f5) {$F_5$};
    \node [block, above= of f2] (c1) {$C_1$};
    \node [block, right= of c1] (c2) {$C_2$};
    \node [block, right= of c2] (c3) {$C_3$};
    \path [line] (c1) -- (f1);
    \path [line] (c1) -- (f2);
    \path [line] (c2) -- (f2);
    \path [line] (c2) -- (f3);
    \path [line] (c2) -- (f4);
    \path [line] (c3) -- (f4);
    \path [line] (c3) -- (f5);
    \path [line] (s1) -- (f1);
    \path [line] (s2) -- (f2);
    \path [line] (s3) -- (f3);
    \path [line] (s4) -- (f4);
    \path [line] (s5) -- (f5);

\end{tikzpicture}
\caption{Example factor graph, containing~three classifiers $C_1\dots C_3$ and~five samples $S_1\dots S_5$. The structure of the factor graph determines the overlap among the classifiers.}
\label{factorgraph}
\squeezeup
\end{figure}

\begin{ex}
Consider the example factor graph in Figure~\ref{factorgraph}.
The graph consists of~8 variables:~3 named $C_1$--$C_3$ representing~3 different classifiers and~5 variables named $S_1$-$S_5$ representing samples.
These variable are connected using~5 factors, $F_1$--$F_5$, such that \mbox{$F_i(\bar{C},S_i) = \vee \bar{C}\rightarrow S_i$}, where $\bar{C}$ is the set of classifiers with edges entering the factor.

The variables $S_i$ are treated as observations, meaning their value is set, and the variables $C_i$ are treated as query variables.
The inference algorithm chooses at which probability is each $C_i$ set to \code{true}, such that the factors satisfy the observations.
If we set all observations $S_i$ to \code{true}, the inference of the factor graph returns the trivial solution of always setting all $C_i$ to \code{true} (meaning $C_i$ is \code{true} with probability~1.0).
When we set some $S_i$ to \code{false}, the inference algorithm is able to provide more elaborate answers.
For example, setting $S_4$ to false, results in probability~1.0 for $C_1$ and probability~0.5 to both $C_2$ and $C_3$.
\end{ex}

Given the sets of enabled and disabled samples from the previous step, we translate the problem to a factor graph as follows:
\begin{enumerate}\itemsep=-1pt
  \item For each classifier $i$ we define a query variable $C_i$;
  \item for each sample $j$ we define an observation variable $S_j$ and a factor $F_j$;
  \item if sample $j$ was set to be enabled we set $S_j$ to \code{true}, otherwise to \code{false};
  \item we connect each $S_j$ to its corresponding $F_j$;
  \item for every pair of classifier and sample $(i,j)$, if classifier $i$ flags sample $j$ we connect $C_i$ to $F_j$.
\end{enumerate}
Using this construction, we get a factor graph similar in structure to the graph in figure~\ref{factorgraph}.
The inferred probabilities for the query variables~$C_i$ are used as the sampling rates for the corresponding classifiers.

To solve the factor graph and infer the probabilities for $C_i,$ we use Microsoft's Infer.NET~\cite{infernet}, a framework for running Bayesian inference in graphical models\footnote{We use \code{ExpectationPropagation} as the inference algorithm, which proved empirically fastest in our experiments.}.
We evaluate the performance of the solver in Section~\ref{sec:expt}.

\subsection{Discussion}

\point{Maintaining the dataset}
We intend to build our dataset using samples flagged as malicious by our classifiers.
This kind of dataset will naturally grow over time to become very large.
Two problems arise from this situation.

The first problem is that after a while most of the dataset will become outdated.
While we usually wouldn't want to completely drop old samples, since they still represent possible attacks, we would like to give precedence to newer samples over older ones (which essentially should result in higher sampling rates for current attacks).
To facilitate this we can assign a weight to each sample in the dataset.
We represent these weights using $W\in[0,1]^{|D|}$ and rewrite the formulas from~\ref{noweights} as:
\begin{subequations}
\label{weights}
\begin{equation}\label{tp_weighted}
TP(\alpha) = \frac{\Sigma_{i=0}^{|D|}(W_i \cdot G_i \cdot Pr_i(\alpha))}{\Sigma_{i=1}^{|D|}(W_i \cdot G_i)}
\end{equation}
\begin{equation}\label{fp_weighted}
FP(\alpha) = \frac{\Sigma_{i=0}^{|D|}(W_i \cdot (1-G_i) \cdot Pr_i(\alpha))}{\Sigma_{i=1}^{|D|}(W_i \cdot G_i)}
\end{equation}
\begin{equation}\label{tn_weighted}
TN(\alpha) = \frac{\Sigma_{i=0}^{|D|}(W_i \cdot (1-G_i) \cdot (1-Pr_i(\alpha)))}{\Sigma_{i=1}^{|D|}(W_i \cdot (1-G_i))}
\end{equation}
\begin{equation}\label{fn_weighted}
FN(\alpha) = \frac{\Sigma_{i=0}^{|D|}(W_i \cdot G_i \cdot (1-Pr_i(\alpha)))}{\Sigma_{i=1}^{|D|}(W_i \cdot (1-G_i))}
\end{equation}
\end{subequations}
While many different weighting techniques can be used, two examples are:
(1)~Assign weight~0 to all old samples, essentially dropping old samples from the dataset;
(2)~Assign some initial weight $w_0$ to each new sample and exponentially decrease the weights of all samples after each sampling rate selection.
%
The second problem stems from the sampling rates themselves. Given 2 classifiers, $C_1$ and $C_2$, and their corresponding sampling rates, $\alpha_1$ and $\alpha_2$, if $\alpha_1$ is higher than $\alpha_2$ the dataset will contain more samples of attacks blocked by $C_1$ than by $C_2$. This creates a biased dataset that, in turn, will influence sampling rates selected in the future.
This problem can also be addressed using the weights mechanism.
One possible approach will be to assign weights in reverse ratio to the sampling rates (so that samples matching $C_2$ will be assigned a higher rate than samples matching $C_1$).
Other viable approaches exist and the most suitable approach should be chosen based on the setting in which the classifiers are used.

\point{Minimum sampling rates}
In Section~\ref{formalizing} we've defined a sampling rate as $\alpha \in [0,1]^{|C|}$. This definition allows for a complete disable of a classifier by setting it's sampling rate to~0.
In practice, since we can never be sure that an attack has completely disappeared from the landscape, it is unlikely that we will want to completely disable a classifier.

A possible approach to address this is by setting a minimal sampling rate for the classifier. Given that the attack for which this classifier was intended is extremely unlikely to be encountered we don't want to apply the classifier to every sample encountered. however since the attack is still possible we should statistically apply the classifier to some samples to maintain some chance of blocking and noticing an attack (if one appears).
Given an inferred sampling rate $S$, the minimal sampling rate can be introduced in many forms, such as
\begin{itemize}\itemsep=-1pt
\item a lower bound $L$ on the sampling rate assigned to each classifier ($S>=L$);
\item a constant value $X$ added to the sampling rate ($S+X$);
\item some percentage $Y$ reduced from the non-sampled portion ($S+(1-S)\cdot Y$).
\end{itemize}
We note that the minimum sampling rate for each classifier should be proportional to the severity of the attacks for which it was intended.
If the impact of a successful attack is minuscule, we may set a lower minimum sampling rate because even if we miss the attack the consequences are not severe.
However, if the impact is drastic, meaning the severity of the attack is high, then we should set a higher minimum sampling rate as a precaution.

We can formalize the notion of minimal sampling rates as $MinSR\in[0,1]^{|C|}$, which is based on a severity mapping $S\in\mathbb{N}^{|C|}$ (such that $S_j$ is the severity of the attacks for which classifier $C_j$ was intended), and use $MinSR_j$ as either $L$,$X$, or $Y$ from the examples above.

%% file: eval.tex
\section{Experimental Evaluation}
\label{sec:expt}
In this section we first describe our simulation design and then discuss both how much our approach helps with achieving optimization objectives such as reducing false positives, and how long it takes to solve the optimization problems on a daily or weekly basis.

\subsection{Simulation Design}
\label{sec:simDesign}

To evaluate the benefits of our approach we performed several simulations mimicking real-world anti-virus activity.
For the purposes of our simulation, we collected detailed information from Snort signature activity summaries from~12/30/2007 until~9/6/2016, entailing signature additions, updates and removals, as shown in Figure~\ref{fig:snort-add-remove}.
In total, we've collected information regarding~\numTotalSignatures signatures.
\hide{, each of which we use as a classifier in our simulation.
In practice, we use~\empirical{3,029} of these signatures in our simulation (see appendix~\ref{sec:appendix} for details).}

\point{Generating simulated malware traffic}
We generate malware traffic traces (observed true positive and false positive samples) based on the collected Snort signature information.
We base these traces on the following assumptions:
\begin{enumerate}[i]\itemsep=-1pt
\item Each signature was introduced to counteract some specific malware.
\item Nonetheless, some signatures might unintentionally flag malware other than the one it was intended for (resulting in classifier overlap).
\item The main purpose of signature updates is to address some false negatives, resulting in increased true positive and false positive observations.
\end{enumerate}

The following design decisions are also incorporated into the trace generation:
\begin{enumerate}[i]\itemsep=-1pt
\item The decline in true positive observations count for a specific signature is modeled as a power law decay curve.
\item False positive observations count is modeled as a percentage of the true positive count (denoted as a simulation parameter $\theta$).
\item Amount of true positive traffic does not affect false positive observations.
\item Observations may be captured by more than one signature (referred to as ``classifier overlap'').
\end{enumerate}
For more details, please see Section~\ref{sec:design-decisions}.

\point{Simulation scenario}
We aim to simulate a real world usage in our simulation.
The scenario we are simulating is when \emph{once every~3 days} our tool is applied to the latest observations and updates the sampling rates for all active signatures.
New signatures might still be introduced between sampling rate updates and are set to full sampling until the next update.
We believe this to be a reasonable setting that is quite likely to be implemented in practice.

Additionally, under some conditions, Infer.NET's inference algorithm might fail.
Such conditions are very rare (inference for only \overlapFailedDaysPercentage of days either failed or timed-out). However, if they occur we allow the simulation to keep using the sampling rates computed on the last update, which we believe to be a solution applicable in practice.

We defined our optimization goal using a budget-aware objective.
Assuming a known estimated cost ratio $\beta$ between false negatives and false positives, we phrase the objective as $FP+\beta\cdot FN$. We leave $\beta$ as a parameter for the simulation.

Essentially, we set our goal to minimize the overall cost incurred by scanning. We note that, as Section~\ref{formalizing} mentions, there are many possible goals. Based on discussions with commercial vendors, we believe the overall incurred cost is a measure likely to be used in practice.
Another reasonable goal is the total cost of scanning. In this setting each signature would have to be associated with a scanning cost the the system would look for a solution that either minimizes it or keeps it below some given threshold.

\subsubsection{Design Decisions}
\label{sec:design-decisions}

The following design decisions are incorporated into the trace generation:

\begin{figure*}[tb]
\centering
\ifpdf
    \includegraphics[width=0.8\textwidth]{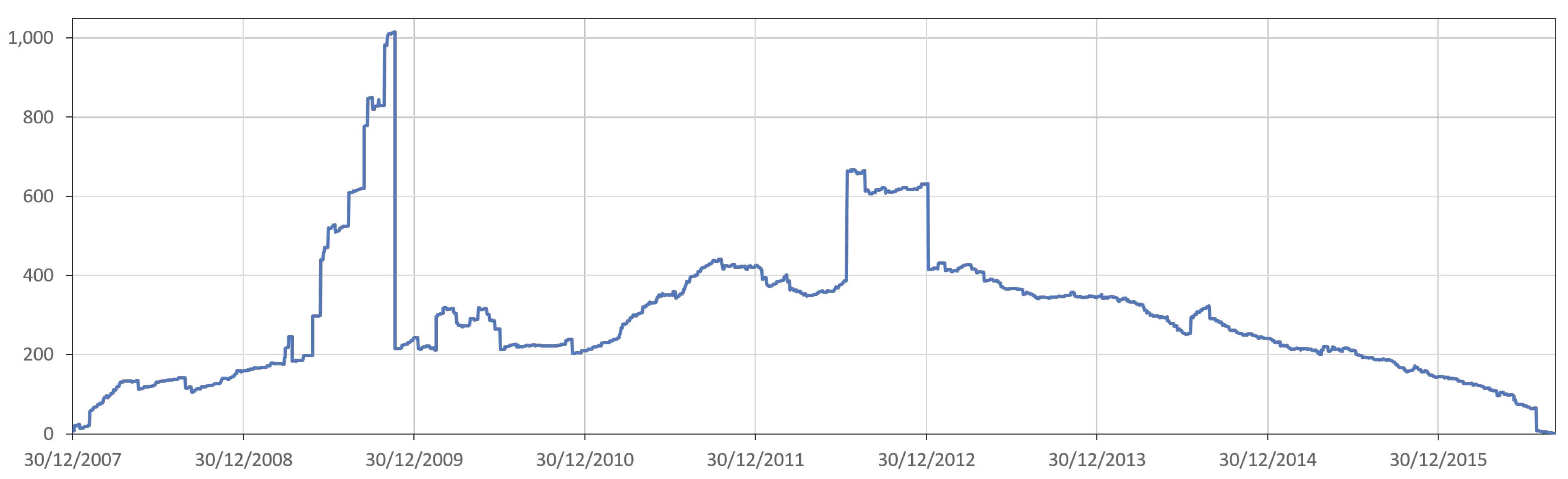}
\fi
\caption{Number of active signatures in the Snort archive (12/30/2007--9/6/2016). Figure~\ref{fig:snort-add-remove} shows some of the update dynamics.}
\label{fig:simAttacks}
\end{figure*}



\point{Simulating the decline in true positives}
We use a power law decay curve to simulate the decline of a specific type of malware over time. Prior research on large-scale exploitation (75~million recorded attacks detected by Symantec sensors) in the wild suggests the power law as an accurate way to model diminishing exploitation rates~\cite{Allodi2015}.

We initialize the curve for an initial true positive count of approximately~\empirical{500} observations per day and calibrate it to fit the lifespan of the signature intended for that malware.
We filter out signatures which were added or removed outside of our sampling period, for which we cannot determine a life span, since we are unable to calibrate a proper decay curve for these signatures.
We also eliminate temporary short-lived signatures (less than~7 days). Out of the~\numTotalSignatures collected signatures, we are left with~\empirical{3,029} signatures after filtering, which we use in the simulation. Figure~\ref{fig:simAttacks} shows the number of active signatures for each day of our simulation.

We note that in many real-world cases, a signature is introduced only a few days after a malware appears in the wild and is removed \emph{at least} a few days after the relevant malware disappeared from the attack landscape. We adjust the overall signature lifespan, introduction and removal dates accordingly to incorporate this insight into the attack model.

\point{Simulating false positives}
We model the amount of false positive observations as a percentage of the true positive observations count. We denote this percentage as $\theta$, used as a parameter for the simulation.

\begin{figure*}
\centering
\ifpdf
    \includegraphics[width=0.85\textwidth]{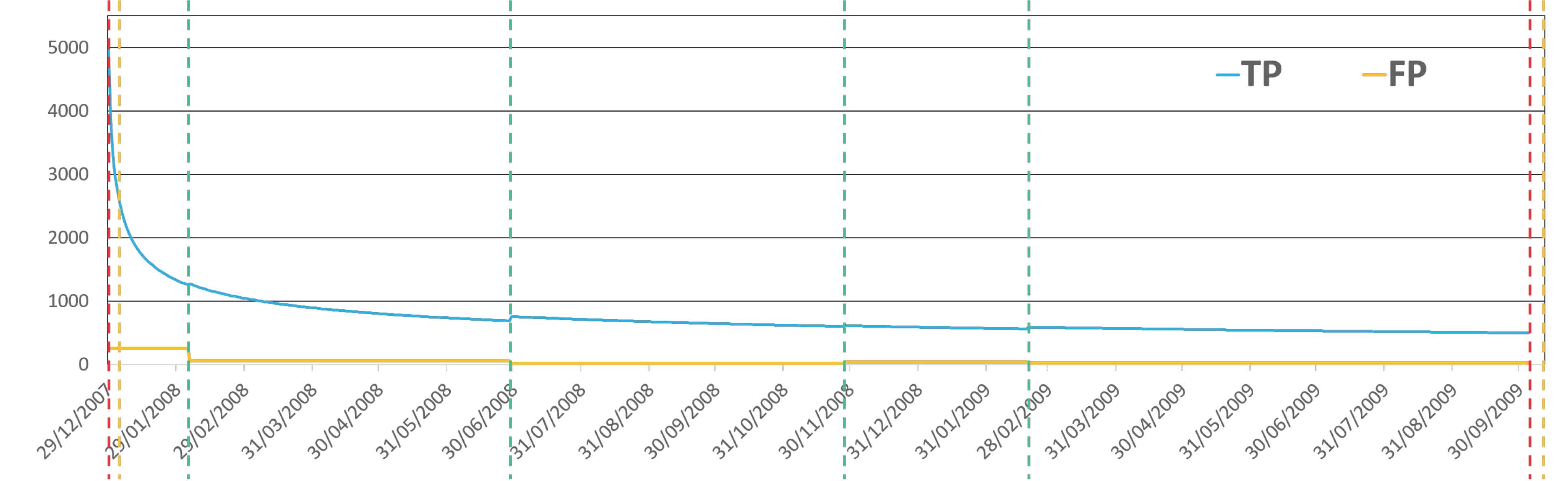}
\fi
\caption{Modeling the true positive count and the false positive count per day for \et signature~2007705, assuming power law decay. The dashed lines at the ends of the figure indicate malware appearance, signature introduction, malware disappearance, and signature removal. The dashed lines in the middle of the figure indicate signature updates.}
\label{fig:modeledAttack}
\end{figure*}

As false positive observations originate solely from legitimate traffic which remains mostly unchanged, we determine that false positive observations should remain constant as long as the signature is not changed. We therefore update the false positive traces only when the signature is updated.
The curve in Figure~\ref{fig:modeledAttack} show an example of the number of true positive and false positive observation for Snort signature~2007705 for a set $\theta$.

\begin{figure}
\centering
\ifpdf
    \includegraphics[width=.8\columnwidth]{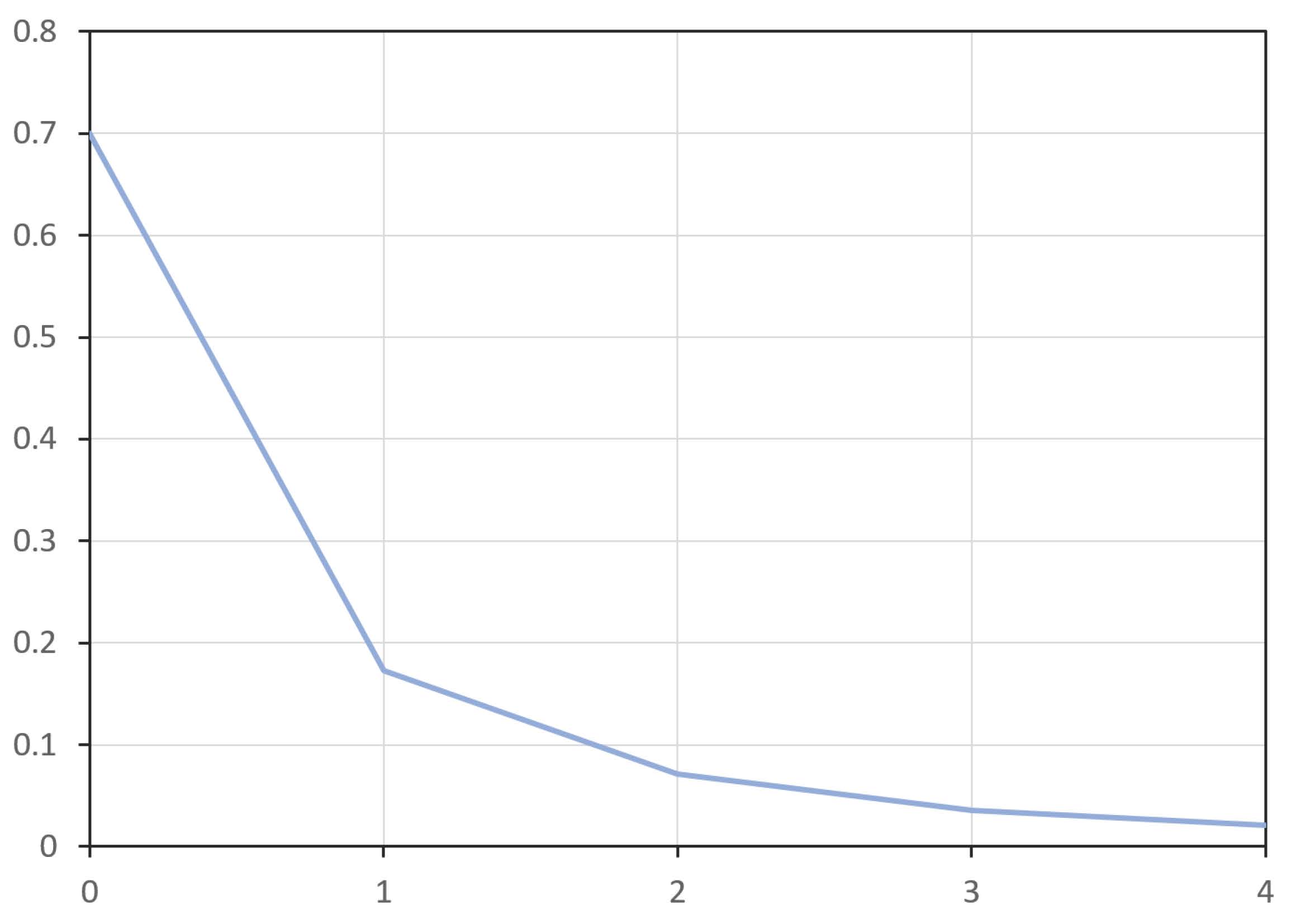}
\fi
\caption{Probability distribution used to decide amount of overlap for each signature.}
\label{fig:overlap}
\squeezeup
\end{figure}

\point{Simulating classifier overlap}
From the collected signature information, we learn that, while there exists some overlap between signatures, most signatures only flag one kind of malware.
To simulate overlap we randomly choose for each signature and each observation with how many other signatures it overlaps.
We draw this value from the distribution shown in Figure~\ref{fig:overlap}, calibrated to match our collected signature information.

\subsection{Experimental Setup}
To compare different simulation conditions, we ran several simulations, each with a different combination of values for $\theta$ and $\beta$, both with classifier overlap and without.
The simulations were executed on a Linux machine with~64 AMD Opteron(TM)~6376 processors, operating at~2.3GHz each, and~128~GB RAM, running Ubuntu~14.04.
Each simulation was assigned the exclusive use of a single core.

\subsection{Precision and Recall Results}

\begin{figure}[tb]
\centering
\ifpdf
\includegraphics[width=.9\columnwidth]{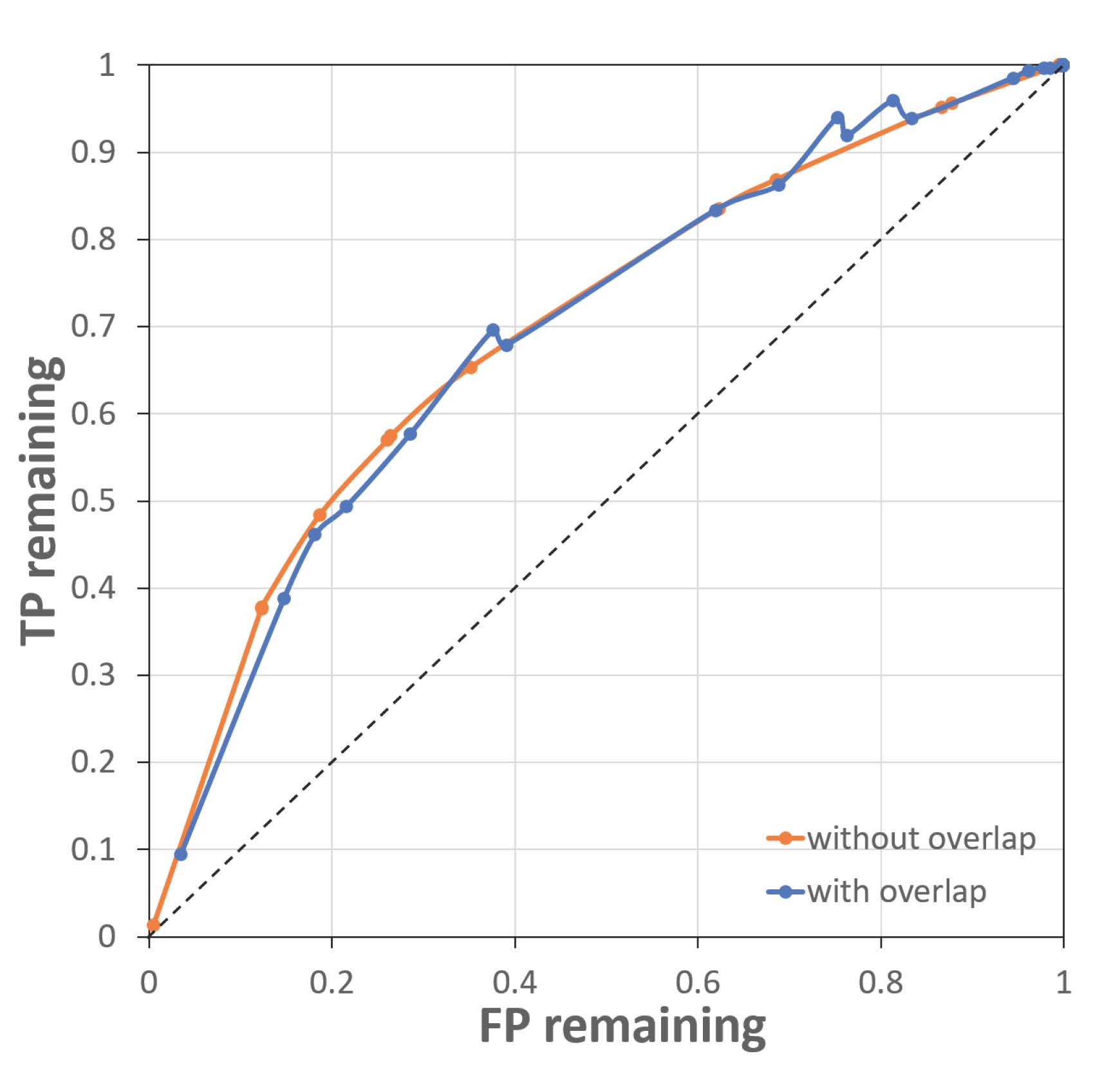}
\fi
\caption{Percentage of true positives remaining compared to percentage of false positives eliminated with and without classifier overlap.
Different setting for FP-threshold and cost ratio correspond to different points on the curves.
}
\label{fig:roc}
\end{figure}

By applying the sampling rates computed by our system, one can eliminate part of the false positives previously observed at the expense of losing part of the true positive observations. In Figure~\ref{fig:roc}, we show the percentage of true positives remaining compared to the percentage of false positives remaining.

The dashed line across each of the figures symbolizes an equal loss of both false positives and true positives.
The area above the dashed line matches settings in which less true positives are lost compared to false positives. This is the area we should strive to be in, since it represents a sampling which is relatively cost-effective.
One can clearly see from the figures that, regardless of classifier overlap, all of our simulations reside above the dashed line.

\begin{figure*}[tb]
\centering
\begin{subfigure}{.25\textwidth}
  \centering
    \ifpdf
        \includegraphics[width=\columnwidth]{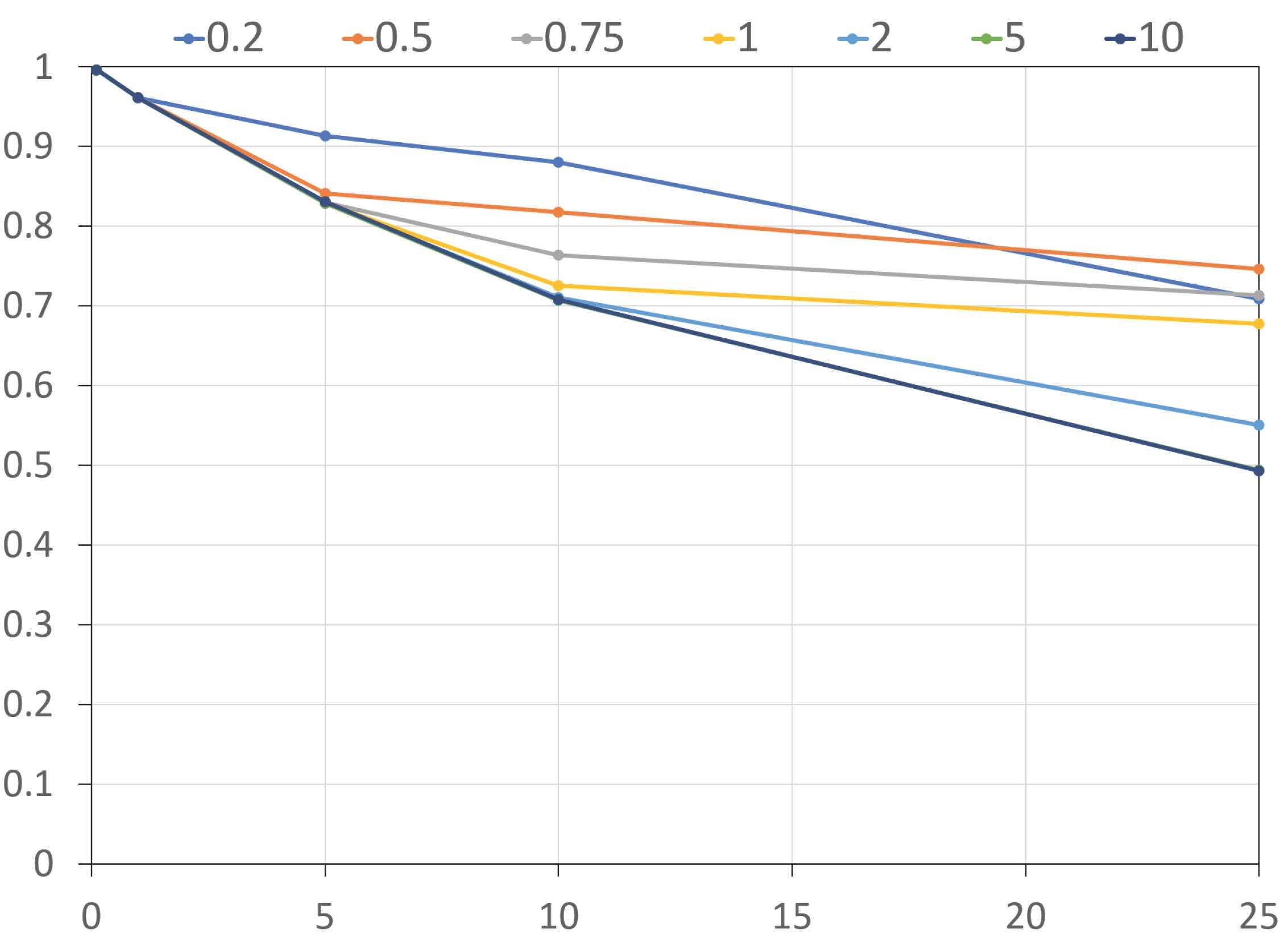}
    \fi
  \caption{precision without overlap}
  \label{fig:precision_01}
\end{subfigure}%
\begin{subfigure}{.25\textwidth}
  \centering
    \ifpdf
        \includegraphics[width=\columnwidth]{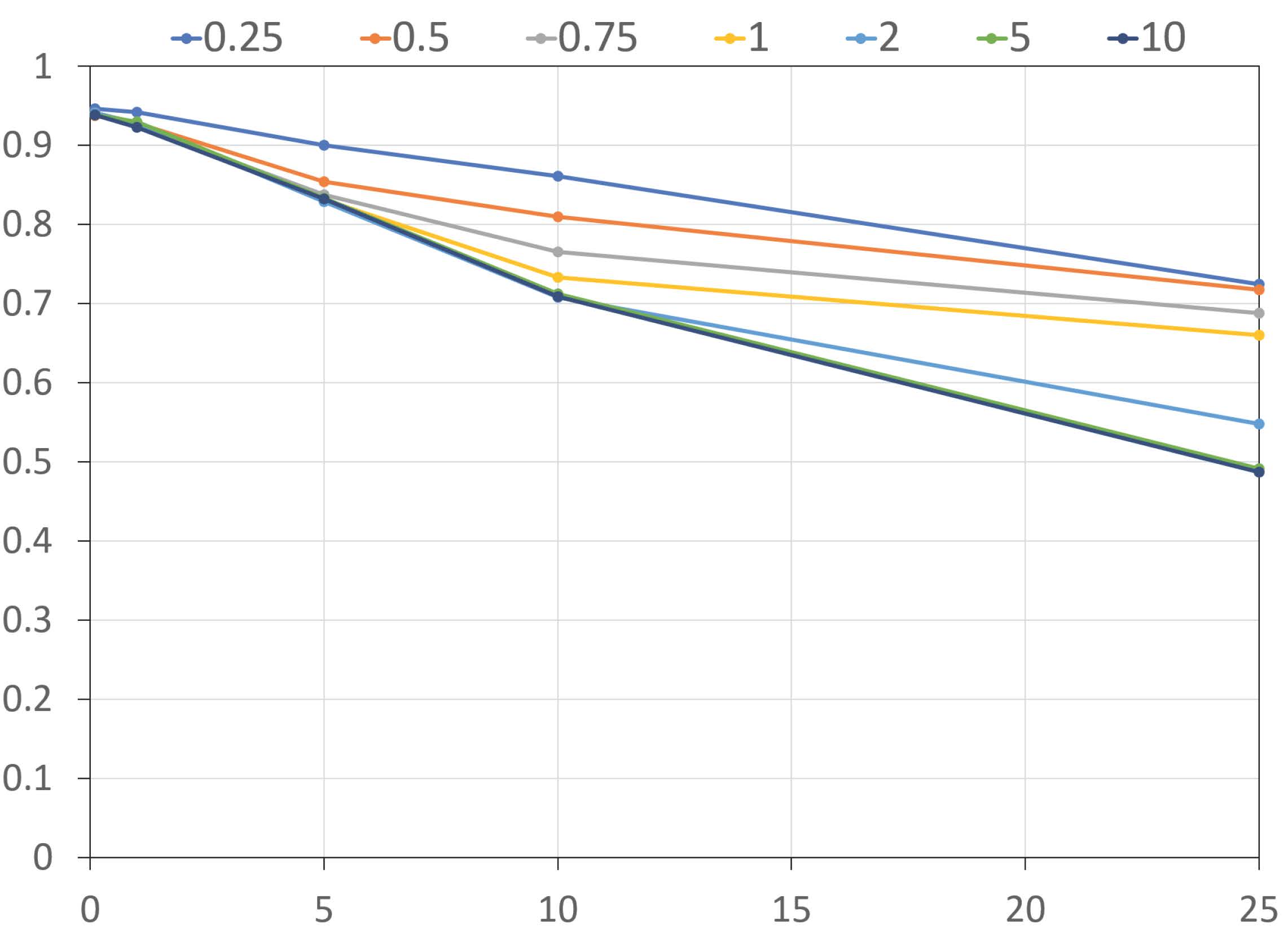}
    \fi
  \caption{precision with overlap}
  \label{fig:precision_sampling}
\end{subfigure}%
\begin{subfigure}{.25\textwidth}
  \centering
    \ifpdf
        \includegraphics[width=\columnwidth]{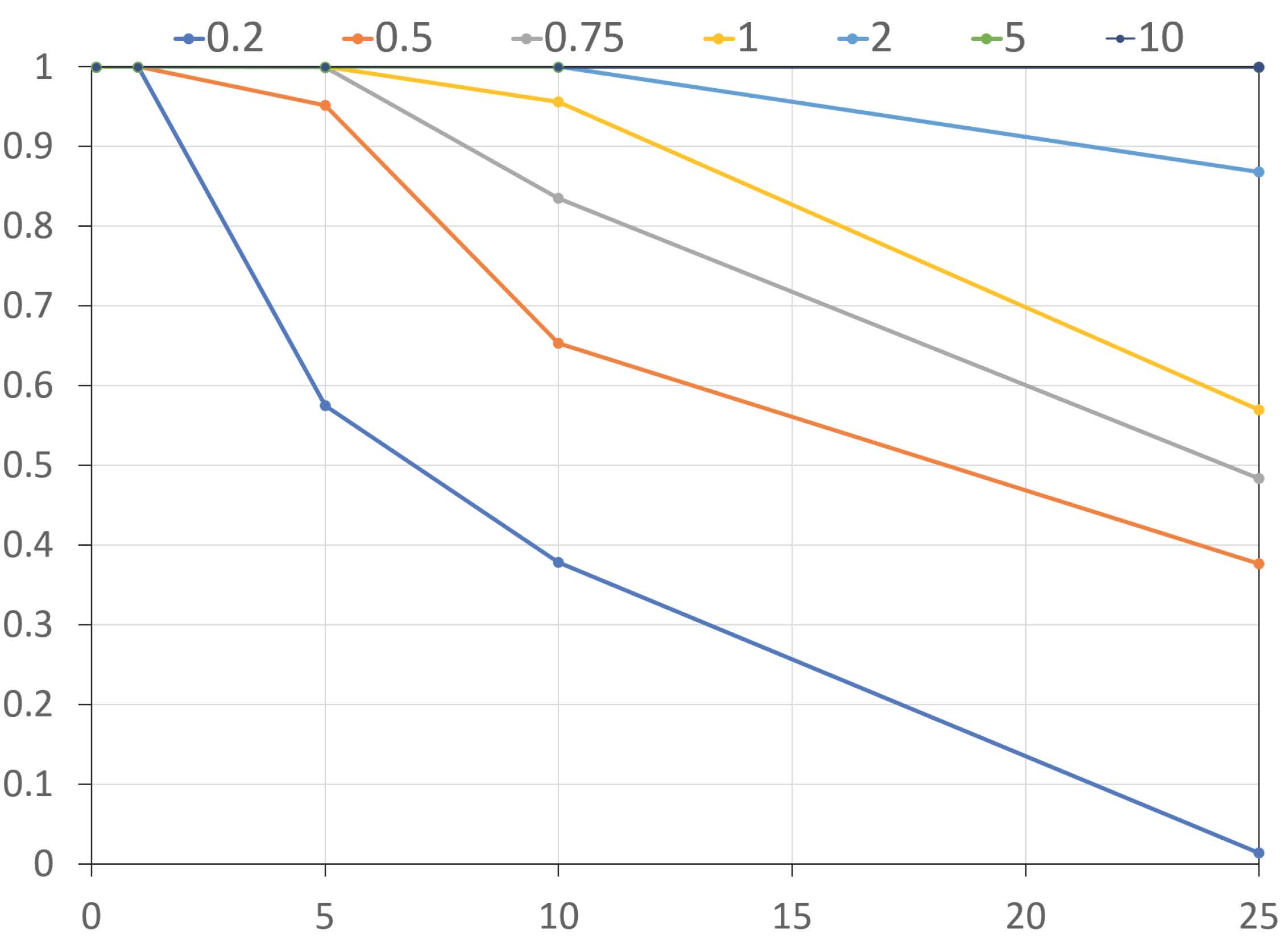}
    \fi
  \caption{recall without overlap}
  \label{fig:recall_01}
\end{subfigure}%
\begin{subfigure}{.25\textwidth}
  \centering
    \ifpdf
        \includegraphics[width=\columnwidth]{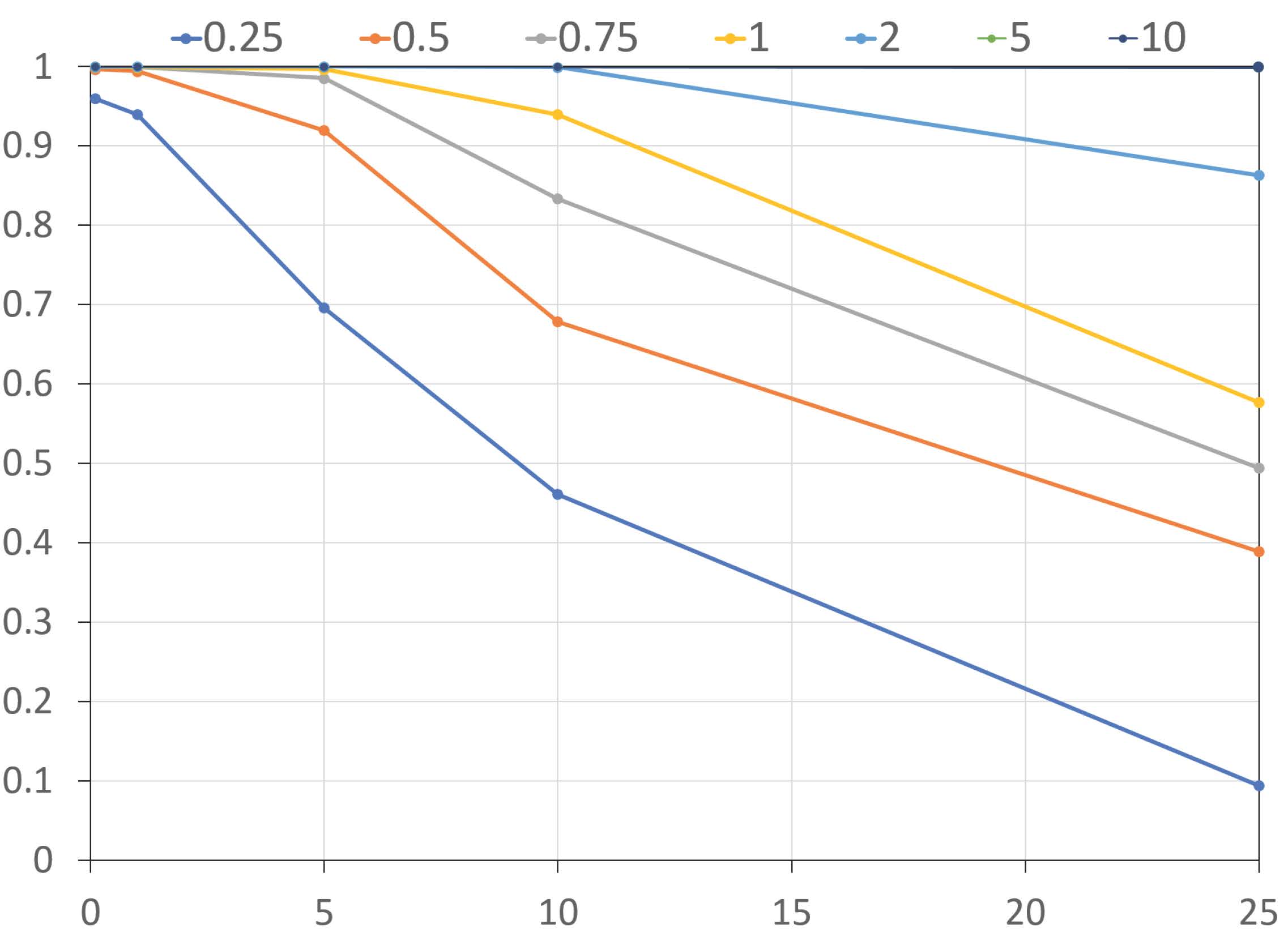}
    \fi
  \caption{recall with overlap}
  \label{fig:recall_sampling}
\end{subfigure}%
\caption{Classification precision and recall as a function of $\theta$ for different values of $\beta$ with and without classifier overlap.}
\label{fig:precision_and_recall}
\squeezeup
\end{figure*}

Figure~\ref{fig:precision_and_recall} show the classification precision and recall as a function of $\theta$ for different values of $\beta$, both with and without classifier overlap.
The figures show that, regardless of overlap, both precision and recall drop when $\beta$ and $\theta$ rise.
A rise of $\theta$ means there are more false positive observations, which reduces the portion of observed true positives, thus affecting the overall precision and recall.
Similarly, a rise of $\beta$ means that relative cost of a false negative is higher than that of a false positive. Therefore, based on the optimization objective we set in Section~\ref{sec:simDesign}, it is only logical that the system will choose to allow for more false positives, rather than risking a false negative, thus again affecting both precision and recall.

\point{Adapting to the situation}
From the aforementioned figures, we learn that the effectiveness of applying sampling rates depends greatly on the operating scenario.
In some cases, where for example false negatives are extremely expensive (as might be the case for corporate datacenters), the sampling rates remain rather high and thus the overall true positive and false positive counts remain mostly unaltered.
On the other hand, when false negatives are relatively cheap (as is often the case for private, user owned desktops), we can expect our system to determine sampling rates that are relatively low.

We believe our system is especially well suited for large-scale services such as web-mail providers. In this scenario, each user can set it's own willingness to accept risk, i.e. the subjective cost of a false negative. Doing so will allow the service's servers to greatly reduce their scanning workload, which will have a tremendous effect on the overall server performance as the amount of data scanned by these servers is huge. Such services would like obtain the greatest benefit from our approach.

\subsection{Solution Times}

We recognize that for a system such as the one proposed in this paper to be applicable to real world scenarios, it is required that solving and computing new sampling rates be very fast and cheap. Long solving times mean the system would not be able to quickly adapt to changing landscapes and respond by setting new sampling rates in a timely fashion.

\begin{figure}[tb]
\centering
\ifpdf
    \includegraphics[width=.9\columnwidth]{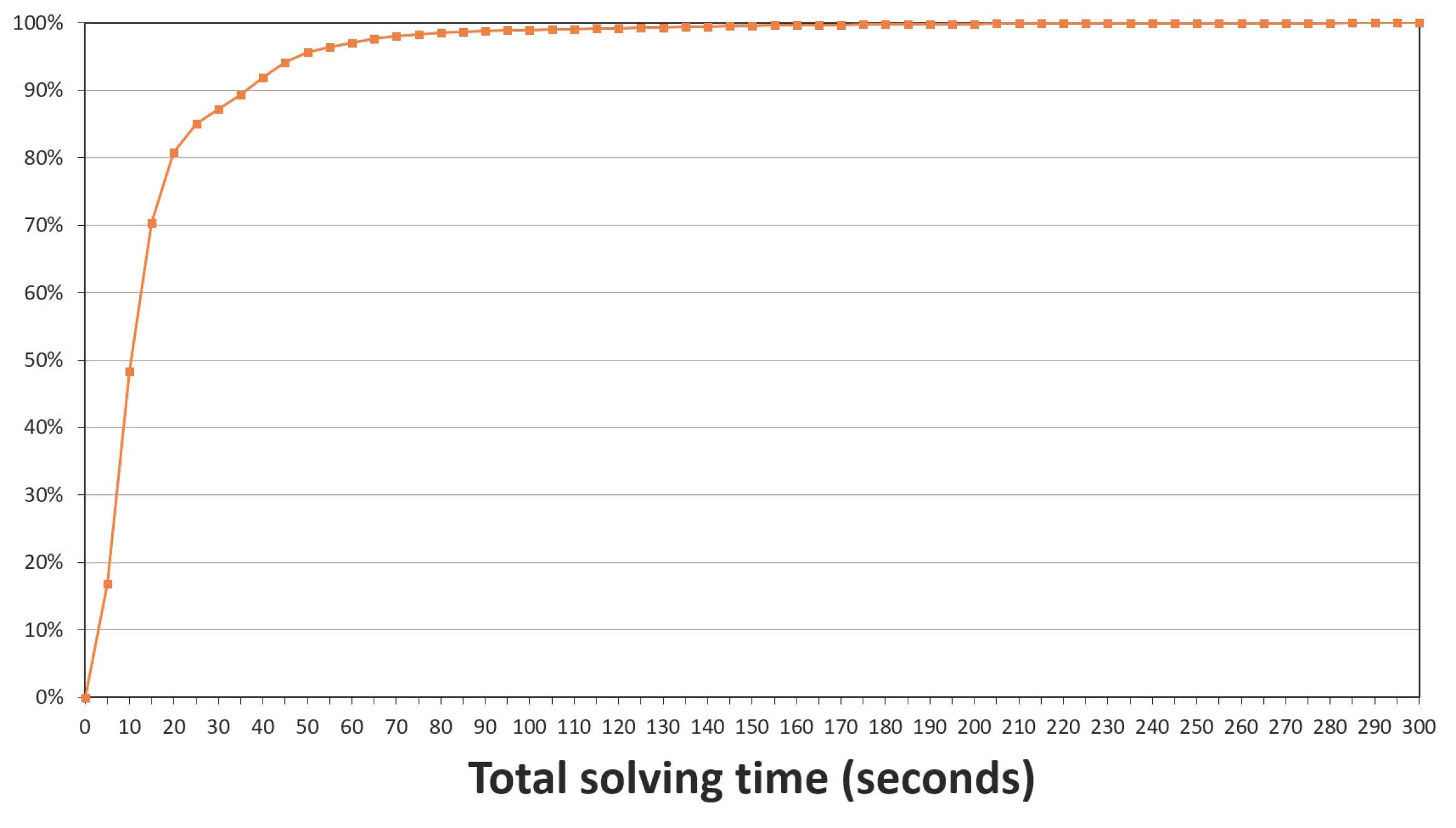}
\fi
\caption{Cumulative distribution of overall solution times (PuLP + Infer.NET).}
\label{fig:times-histogram}
\end{figure}

Figure~\ref{fig:times-histogram} shows a cumulative distribution of the total solving times (both PuLP and Infer.NET) measured during our simulations for each day.
The figure shows that over~\empirical{80\%} of simulated days were solved in under~\empirical{20} seconds.
The day which took our system the longest to solve took less than~\empirical{5} minutes~(\empirical{285} seconds to be exact).
This tells us that using this kind of system in a responsive manner is indeed feasible.

\begin{figure}[tb]
\centering
\ifpdf
    \includegraphics[width=.9\columnwidth]{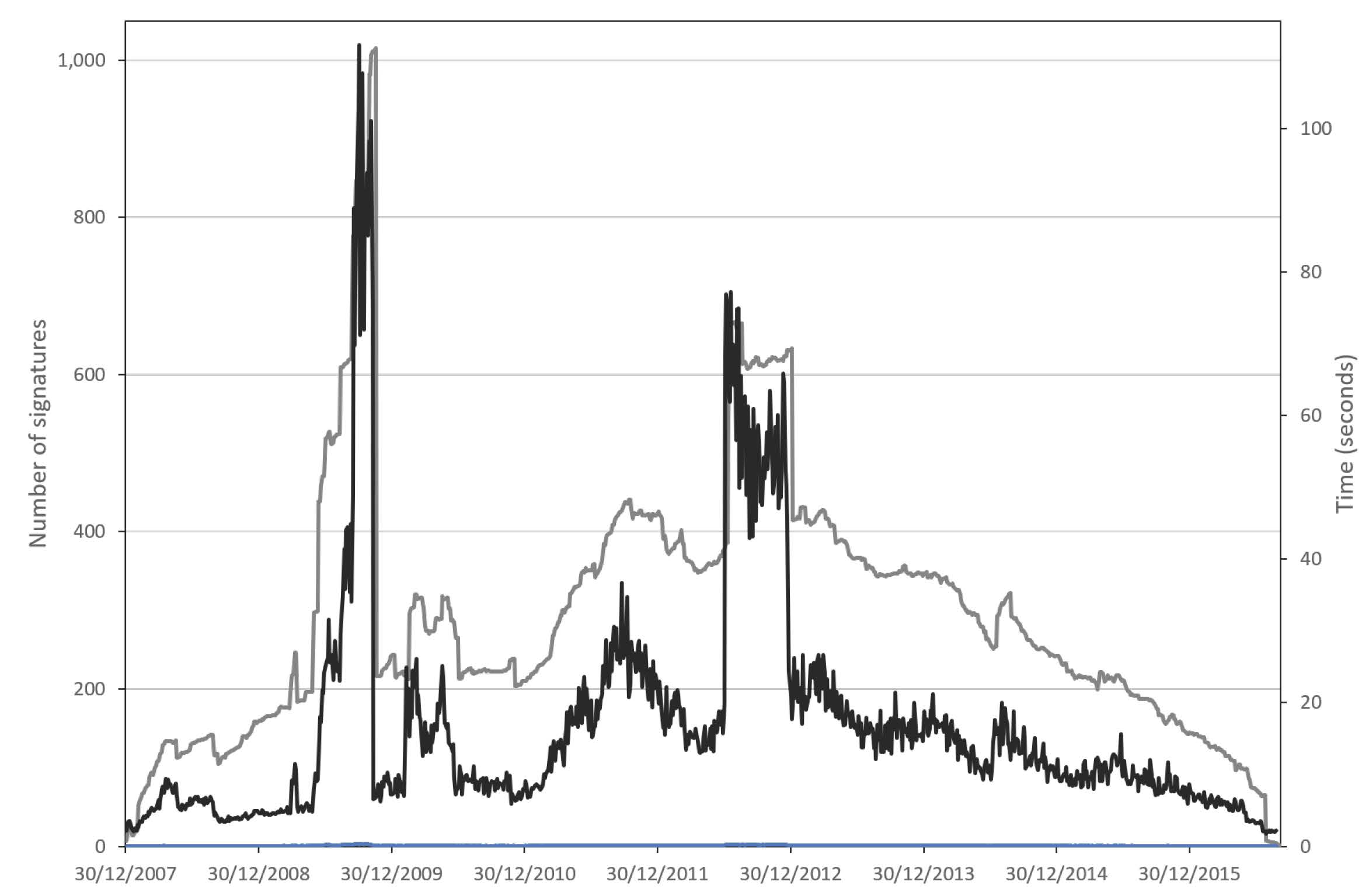}
\fi
\caption{Average daily solving times.
The blue line (at the bottom) represents PuLP solving time (\empirical{$\le 1$} second).
The black line represent Infer.NET solving time.
The grey line in the background show number of active signatures per day. }
\label{fig:solution-times}
\squeezeup
\end{figure}

Figure~\ref{fig:solution-times} shows the average solving time needed for each day of our simulation.
We first note that the solving times for PuLP (represented by the blue line) are extremely low, constantly below 1 second. When there is no classifier overlap, the solution provided by PuLP is sufficient as the sampling rates for the classifiers. This means that when there is no overlap, solving is extremely fast.
%
Also, there is a clear correlation between Infer.NET solving time and the number of active signatures, both following the same trends. This correlation is interesting as it indicates that
(1)~solving time can be anticipated in advance, and
(2)~we can accelerate the solving of days with many active signatures using a ``divide-and-conquer'' approach, meaning we can split them to smaller batches and solve each one separately. 
%


\subsection{Experimental Summary}

\point{Reduction in false positives}
Regardless of classifier overlap, when comparing the reduction in the number of false positives to that of true positives, we find that our responsive optimization technique removes \emph{more} false positives than true positives, both in terms of percentages (\empirical{19.23\%} compared to~\empirical{12.39\%} without overlap;~\overlapFpReductonPercentage compared to~\overlapTpReductonPercentage with overlap) with overlap) and in terms of absolute values (\empirical{9,286,530} compared to~\empirical{8,002,871.5} without overlap;~\empirical{9,225,422.6} compared to~\empirical{8,065,888.6} with overlap). The reduction in absolute values is surprisingly significant,  considering that the highest value of $\theta$ in our simulations was set to~\empirical{25\%} of the true positive rate.
In settings with classifier overlap, applying sampling rates is \emph{more beneficial} than in settings without classifier overlap. This can be observed from the relevant reduction rates,~\overlapFpReductonPercentage compared to~\empirical{19.23\%} of false positives and~\overlapTpReductonPercentage compared to~\empirical{12.39\%} of true positives. This means that, on average, we can eliminate more false positives at the expense of fewer true positives.

While in some settings the benefits of our approach are not as drastic as in others, our simulations clearly demonstrate that this approach is advantageous regardless of the operating scenario.

\point{Solver running time}
In settings \emph{without} classifier overlap, the sampling rates for all simulated days were computed in mere seconds.
In settings with overlap, timing measurements indicate that our approach for setting sampling rates is computationally feasible and applicable to real-world settings. The sampling rates for over~\overlapSolvedDaysPercentage of simulated days were computed in under~\overlapSolvedDaysTime minutes per day, with an average and median of~\empirical{15} seconds. 

%% file: limitations.tex
\section{Limitations}
\label{sec:limitations}

We acknowledge that our approach has the following limitations. 

\hide{
\point{Reliance on simulation}
In this paper, we evaluate our approach based solely on simulation. Doing so allows us to compare the relative gains our approach provides compared to existing practices of handling outdated signatures.
	
We did not evaluate using a real-world experimental data because we could not obtain access to a large enough daily dataset of flagged samples needed to train and adapt our technique. Such a dataset is likely only available to commercial security vendors such as Symantec of McAfee, although it is far from clear that these types of vendors are likely to provide access to the data we would need. 
}
	
	\hide{
A simulation also has the added advantage of allowing us to verify that the benefits we predicted are actually observable by using our approach. We did not wish to endanger any users involved in a real-world experiment by leaving them exposed to attacks without any practical guarantees.}

\point{Code complexity}
The adoptive approach may increase development and maintenance costs to developers or security researchers, because a new mechanism for introducing the sampling rates will need to be implemented.
We also note that debugging might prove more difficult since, in addition to the inputs introduced to the system, the sampling rates used would also have to be taken into account, leading to possible non-determinism.
	
When it comes to the issue of long-term cost of enforcement, we remind the reader that our approach doesn't \emph{remove} signatures, it merely disables them.
	
\point{Adaptive attacker countermeasures}
Most security mechanism entail new opportunities for attackers and our approach is no exception. An attacker familiar with the working of a system utilizing our approach could attempt to use it to her advantage. With the approach in this paper, classifiers are no longer always applied and any attack has some probability to get through the classifiers if it is not sampled. Using this knowledge, the attacker can attempt the same attack several times, hoping that at least one instance will be overlooked.

Additionally, classifiers for uncommon attacks would be sampled at lower rates. An attacker can deliberately use \emph{outdated} vulnerabilities, which are likely not sampled frequently to increase her chances of a successful attack. 
The attacker can use a variety of attacks to force a high sampling rate on \emph{all} classifiers. This would at most cause the system to revert to the current default situation, in which all classifiers are always on.
	
Using one or more of the approaches outlined above, the attacker can \emph{temporarily} increase her chances of success. Doing so will increase the amount of observed attacks for the relevant classifiers, thus triggering an increase in respective sampling. As a result, we anticipate that the window of vulnerability will be small, and that relatively few users will be affected.

	%

%% file: related.tex
\section{Related Work}
\label{sec:related}

\point{Reactive vs. proactive Security}
There has been some interest in comparing the relative performance of proactive and reactive security over the years.
Barth~\etal~\cite{Barth:FC:2010} make the interesting observation that proactive security is not always more cost-effective than reactive security.
Just like in our paper, they support their claim using simulations.
%
Barreto~\etal~\cite{barreto2013controllability} formulate a detailed adversary model that considers different levels of privilege for the attacker such as read and write access to information flows.

Blum~\etal~\cite{Blum:2014} demonstrate that optimizing defences as if attackers have unlimited knowledge is close to optimal when confronted with a diligent attacker, with limited knowledge regarding current defences prior to attacking.
Such attackers are more realistic, thus supporting our results highlighting the benefits of tunable security.

Classifiers designed to detect malicious traffic or samples are often affected by an issue known as ``Concept drift''~\cite{Widmer1996}. Researchers have proposed a number of ways to address concept drift through retraining classifiers. For instance, Soska and Christin~\cite{DBLP:conf/uss/2014} 
present techniques that allow to proactively identify likely targets for attackers as well as sites that may be hosted by malicious users.

Several researchers have proposed generating new signatures automatically~\cite{newsome2005polygraph,perdisci2010behavioral,griffin2009automatic,sathyanarayan2008signature,zolkipli2010framework}. Signature \emph{addition} seems to largely remain a manual process, supplemented with testing potential AV signatures against known deployments, often within virtual machines. Automatic signature generation will likely improve security but will worsen the false positive and performance issues addressed in this paper.

\point{Exploitation In the Wild}
Nayak~\etal~\cite{exploited-in-wild-raid} highlights the lack of a clear connection between vulnerabilities and metrics such as attack surface or amount of exploitation that takes place. They use field data to get a more comprehensive picture of the exploitation landscape as it is changing. None of the products in their study had more than~35\% of their disclosed vulnerabilities exploited in the wild. 
These findings resonate with the premise of our paper.

One of the better ways to understand the exploitation landscape is by consulting
intelligence reports published by large security software vendors. Of these, reports published by Microsoft~\cite{microsoft-intelligence-20} and Symantec~\cite{symantec-2016} stand out, both published on a regular basis. 
A recent report from Microsoft~\cite{windows-10-mitigations,rsa-microsoft-2015} highlights the importance of focusing on exploitation and not only vulnerabilities.
The current approach to software security at Microsoft is driven by data.
This approach involves proactive monitoring and analysis of exploits found in-the-wild to better understand the types of vulnerabilities being exploited and exploitation techniques being used.

\hide{
Sabottke~\etal focus on determining which vulnerabilities are likely to be exploited
after a disclosure~\cite{sabottke2015vulnerability}.
They conduct a quantitative and qualitative
exploration of the vulnerability-related
information disseminated on Twitter. 
}

Bilge~\etal~\cite{Bilge:2012:BWK:2382196.2382284} focus on
the prevalence and exploitation of zero-days.
\hide{Searching this dataset for malicious files that exploit known vulnerabilities indicates which files appeared on the Internet before the corresponding vulnerabilities were disclosed. They identify~18 vulnerabilities exploited before disclosure, of which~11 were not previously known to have been employed in zero-day attacks.}
They discover that a typical zero-day attack lasts~312
days on average and that, after vulnerabilities are disclosed
publicly, the volume of attacks exploiting them increases by
up to~5 orders of magnitude. These findings were important
in designing credible simulations in this paper.

\point{Economics of Security Attacks and Defenses}
Herley~\etal~\cite{Herley:PNAS:2016} point out that, while it can be claimed that some security mechanism improves security, it is impossible to prove that a mechanism is necessary or sufficient for security, meaning there is no other way to prevent an attack or that no other mechanism is needed.
They also make a similar observation stating that one can never prove that a security mechanism is redundant and not needed.
These observations put into words the frame of mind that resulted in the current overwhelming number of active security mechanisms.
We try to address this problem using the proposed sampling rates.

A report by the Ponemon Institute~\cite{Ponemon:2015} estimated the costs of false positives to industry companies. The estimation was based on a survey filled by~18,750 people in various positions.
While the numbers portrayed in the report are not accurate, they do paint an interesting picture.
The average cost of false positives to a company was estimated at~1.27 million dollars per year.
These estimations include the cost analyzing and investigating false positive reports as well as the cost of not responding in time to other true positive reports.

\point{Models of Malware Propagation}
Arbaugh~\etal~\cite{Arbaugh:IEEE:2000} introduced a vulnerability life-cycle model supported by case studies.
The introduced model is different than the intuitive model one would imagine a vulnerability follows.
We relied on the insights presented in this paper in designing our models for trace generation.

Many works have focused on modeling malware propagation.
Bose~\etal~\cite{bose2013agent} study several aspects crucial to the problem, such as user interactions, network structure and network coordination of malware (e.g. botnets).
Gao~\etal~\cite{mobile-malware} study and model the propagation of mobile viruses through Bluetooth and SMS using a two-layer network model. 
Fleizach~\etal~\cite{fleizach2007can} evaluate the effects of malware self-propagating in mobile phone networks using communication services. 

Garetto~\etal~\cite{garetto2003modeling} present analytical techniques to better understand malware behavior.
They develop a modeling methodology based on Interactive
Markov Chains that captures many aspects of the problem, especially the impact of the underlying topology on the spreading characteristics of malware.
%
Edwards~\etal~\cite{edwards2012beyond} present a simple Markov model of malware spread through large populations of websites and studies the effect of two interventions that might be deployed by a search provider: blacklisting and depreferencing, a generalization of blacklisting in which a website's ranking is decreased by a fixed percentage each time period the site remains infected. 

Grottke~\etal~\cite{grottke2015wap} define metrics and models for the assessment of coordinated massive malware campaigns targeting critical infrastructure sectors.
Cova~\etal~\cite{Cova2010} offer the first broad analysis of the infrastructure underpinning the distribution of rogue security software by tracking~6,500 malicious domains.
%
Hang~\etal~\cite{Hang2016InfectmenotAU} conduct an extensive study of malware distribution and follow a website-centric and user-centric point of view. 
%
Kwon~\etal~\cite{tanaka2016analysis} analyzed approximately~43,000 malware download URLs over a period of over~1.5 years, in which they studied the URLs' long-term behavior.
%

\hide{
\subsection{mostly irrelevant}
\point{Evolution of Exploit Kits (2015)}
\cite{TrendMicro:2015}
A report by Trend Micro regarding exploit kits.
They briefly mention some trends in which exploits are being used in kits.
Not enough data to actually base anything of it and not very interesting/related.

\point{Time-To-Compromise Model For Cyber Risk Reduction Estimation (2005)}
\cite{McQueen:2006}
The authors try to estimate how long it will take an attacker to find compromise a system (including finding a new vulnerability, creating an exploit, etc.).
They do not have any real data to support their claims.
I find this paper mostly useless and meaningless to us (here only for documentation).

\subsection{Bypasses}
\point{Scriptless Attacks - Stealing the Pie Without Touching the Sill
(2012)}
\cite{Heiderich:CCS:2012}
This paper examines the attack surface remaining after deployment of xss filters and other mitigation techniques.
They discuss several attacks using CSS, svg, fonts, etc.

\point{Heat-seeking Honeypots: Design and Experience (2011)}
\cite{John:WWW:2011}
Used honeypots to attract attackers.
Couldn't get access to their data.
According to the published results they only found a very small amount of XSS (which I assume is most likely only testers, not real exploitation).

\point{mXSS Attacks: Attacking well-secured Web-Application by using innerHTML Mutations
(2013)}
\cite{Heiderich:CCS:2013}
Bypasses XSS filter based on differences between filtering and rendering.
Uses malformed code injected code that will bypass filter, but will execute after rendering engine (which is assumed to be more permissive) fixes malformed parts.
Applicable only to DOM-based XSS exploited via innerHTML and similar methods.

WAF bypasses: \cite{waf-bypass-bh-2016}
and \cite{protocol-level-evasion-2012}

\point{25 Million Flows Later - Large-scale Detection of DOM-based XSS
(2013)}
\cite{Lekies:CCS:2013}
Used taint analysis in the browser to detect DOM-based XSS vulnerabilities.
Implemented automatic exploit generation to verify found vulnerabilities (focusing on required context escaping, doesn't try to avoid filters)
Taint analysis isn't very scalable or suitable for widespread deployment.

\point{Evaluating the customer journey of crypto-ransomware and the paradox behind it
(2016)}
\cite{FSecure:2016}
Interesting but not really relevant to us.
In the beginning they mention the decline of other attack methods to make room for new, more profitable, attacks (they don't mention XSS specifically, but we can claim that the same holds for XSS which lost prominence because of problems monetizing it).

\point{A Survey on Transfer Learning
(2010)}
\cite{Pan:IEEE:2010}
Transfer learning are methods for applying a model learned use a dataset X to some other dataset Y.
They assume that the dataset characteristics (such as the probability distribution of signals) is known.
Our setting is most similar to the one they refer to as transductive transfer learning, in which the both datasets use the same alphabet and set of labels, X is somewhat labeled, Y is completely unlabeled and the probability distributions of X and Y are different.

This is not directly applicable because of the assumption that Y's probabilities are known but it can be used to updating some learned model for filtering when the difference in probabilities measured over distinct time periods is very small (for example, collect probabilities for each week, if the difference from last week is below some threshold update the model, otherwise retrain).

\point{Formulating Cyber-Security as Convex Optimization Problems
(2013)}
\cite{Vamvoudakis:2013}
Defines the attacker's point-of-view of cyber-security as an optimization problem where the attacker wants to optimize it's reward compared to the resources invested in the attack.
Theory evaluated over logs of attacks from the 2011 iCTF competition.
What we want to do is provide a similar perspective from the defender's point-of-view.

\point{Multi-objective Optimizations}
\cite{Caramia:2008}
This paper summarizes the problem of multi-objective optimizations and the possible solution approaches to it.
From the continuous solution techniques proposed in 2.3, I think:
	- 2.3.3 is the closest to our goals (so far)
	- 2.3.2 and 2.3.4 are other reasonable solutions
	- 2.3.1 is, so far, my least favorite solution

\point{AVCLASS: A Tool for Massive Malware Labeling (2016)}
\cite{Sebastian:RAID:2016}
Presents a large dataset of malware that might be useful for our simulation/evaluation.

\subsection{Specific Attack Techniques}
\point{DieHard: Probabilistic Memory Safety for Unsafe Languages
(2006)}
\cite{Berger:PLDI:2006}
Uses redundancy (and some randomization) to secure against memory-related bugs.
Basic idea is that if you have several replicas of the memory/program and they disagree on some values then that indicates an attacks.

\point{Regular Expressions Considered Harmful in Client-Side XSS Filters
(2010)}
\cite{Bates:WWW:2010}
Explains XSS and the design of XSS filters.
Describes problems with IE filter architecture, emphasizing the fact that the filter parses data separately from rendering engine.
XSSAuditor is a result of this work.

\point{The Conundrum of Declarative Security HTTP Response Headers: Lessons Learned
(2010)}
\cite{Sood:CollSec:2010}
Describes the current state and problems of declarative security (such as CSP).
Problem is that each developer/administrator needs to manually declare these directives for their site.
Security is not applied by default and are therefore not widely deployed yet.

\point{Back in Black: Towards Formal, Black box Analysis of Sanitizers and Filters
(2016)}
\cite{Argyros:IEEESEC:2016}
Presents a method for automatically deducing transducers that represent simple sanitizers (such as HTML and URL encoders/decoders, etc).
This is an interesting idea that might also be applicable in our setting to learn the input/output relation of the server side.
I think for our need they're implementation is overly complicated and a simpler, more naïve, approach can be used.

\point{A tale of the weaknesses of current client-side XSS filtering
(2013)}
\cite{Lekies:BlackHat:2014}
Lists problems/bypasses of XSSAuditor.
Seems most of these problems lie in areas which the XSSAuditor doesn't try to filter or deal with (such as DOM-based XSS , double injections, etc).
Provides a good explanation of how the XSSAuditor filter works.
} 

%% file: conclusions.tex
\section{Conclusions}
\label{sec:conc}

Securing today's complex systems does not come free of charge. The most common cost is performance. 
Using three security mechanism examples, anti-virus signatures, Snort malware signatures, and ad-blocking lists, we show that the cost of security enforcement (measured in terms of latency) often grow linearly with the number of policies that are involved. It is therefore imperative to devise ways to limit the enforcement cost.

In this paper we argued for a new kind of \emph{tunable framework}, on which
to base security mechanisms. This new framework enables a
more reactive approach to security, thus allowing us to optimize
the deployment of security mechanisms based on the current
state of attacks. 

Based on actual evidence of exploitation
collected from the field, our framework can choose which
mechanisms to enable/disable so that we can minimize the
overall costs and false positive rates, while maintaining a
satisfactory level of security in the system.

Our responsive strategy is both computationally affordable and results in significant \emph{reductions} in false positives, at the cost of introducing a moderate number of false negatives.
Through measurements performed in the context of large-scale simulations, we find that the time to find the optimal sampling strategy is mere seconds for the non-overlap case,  and under~\overlapSolvedDaysTime minutes in~\overlapSolvedDaysPercentage of overlap cases.
The reduction in the number of false positives is significant~(about~\empirical{9.2 million}, when removed from traces that are about~\yearsData years long,~\overlapFpReductonPercentage and~\empirical{19.23\%}, with and without overlap,  respectively). 